\documentclass{IEEEtaes}
\usepackage{amsmath,amsfonts}
\usepackage{algorithmic}
\usepackage{algorithm}
\usepackage{lineno,hyperref}
\usepackage{multirow}
\usepackage{epstopdf}
\usepackage{graphicx}
\usepackage{float}
\usepackage{caption}
\usepackage{stfloats}
\usepackage{changes}
\usepackage{color,array,amsthm}
\usepackage{amssymb}
\usepackage{graphicx}
\usepackage{eucal}
\usepackage{mathrsfs}
\usepackage{hyperref}      % include this line if your
\jvol{XX}
\jnum{XX}
\jmonth{XXXXX}
\paper{1234567}
\pubyear{2020}
\doiinfo{TAES.2020.Doi Number}

\newtheorem{theorem}{Theorem}
\newtheorem{corollary}{Corollary}
\newtheorem{lemma}{Lemma}
\theoremstyle{definition}
\newtheorem{definition}{Definition}
\theoremstyle{plain}
\newtheorem{remark}{Remark}
\newtheorem{assumption}{Assumption}

\usepackage[inkscapelatex=false]{svg}
\usepackage[numbers,sort&compress]{natbib}
\usepackage{soul} % 导入 soul 包
\usepackage{color, xcolor} % 颜色包，color 必须导入，xcolor 建议导入

\soulregister{\cite}7 % 注册\cite命令
\soulregister{\citep}7 % 注册\citep命令
\soulregister{\citet}7 % 注册\citet命令
\soulregister{\ref}7 % 注册\ref命令
\soulregister{\pageref}7 % 注册\pageref命令
\modulolinenumbers[5]

\begin{document}

	\title{Adaptive Compatible Performance Control for Spacecraft Attitude Control under Motion Constraints with Guaranteed Accuracy}

	\author{Jiakun Lei}
	\affil{Zhejiang University, China} 
	
		\author{Tao Meng}
		\affil{Zhejiang University, China\\ Hainan Research Institute of Zhejiang University, China} 
		
			\author{Yang Zhu}
		\affil{Zhejiang University, China}
		
		\author{Kun Wang}
		\affil{Zhejiang University, China}
		
		\author{Weijia Wang}
		\affil{Zhejiang University, China}
	%	
	%	\author{Shujian Sun}
	%	\affil{Zhejiang University, China}

	\authoraddress{Jiakun Lei, Kun Wang, Weijia Wang are with the School of Aeronautics and Astronautics, Zhejiang University, Hangzhou 310027, China (Email: leijiakun@zju.edu.cn; wang\_kun@zju.edu.cn; weijiawang@zju.edu.cn). Tao Meng(Corresponding Author) is with the School of Aeronautics and Astronautics, Zhejiang University, Hangzhou 310027, China, and Hainan Reserach Institute of Zhejiang University, Sanya, 572025, China (Email: mengtao@zju.edu.cn). Yang Zhu is with the College of Control Science and Engineering, Zhejiang University, Hangzhou, 310027, China (Email: zhuyang88@zju.edu.cn).}

	\markboth{Lei ET AL.}{SAPPC of Attitude Tracking}
	\maketitle

	\begin{abstract}
This paper focuses on the problem of spacecraft attitude control in the presence of time-varying parameter uncertainties and multiple constraints, accounting for angular velocity limitation, performance requirements, and input saturation.
To tackle this problem, we propose a modified framework called Compatible Performance Control (CPC), which integrates the Prescribed Performance Control (PPC) scheme with a contradiction detection and alleviation strategy. Firstly, by introducing the Zeroing Barrier Function (ZBF) concept, we propose a detection strategy to yield judgment on the compatibility between the angular velocity constraint and the performance envelope constraint. Subsequently, we propose a projection operator-governed dynamical system with a varying upper bound to generate an appropriate bounded performance envelope-modification signal if a contradiction exists, thereby alleviating the contradiction and promoting compatibility within the system.
Next, a dynamical filter technique is introduced to construct a bounded reference velocity signal to address the angular velocity limitation.
Furthermore, we employ a time-varying gain technique to address the challenge posed by time-varying parameter uncertainties, further developing an adaptive strategy that exhibits robustness on disturbance rejection.
By utilizing the proposed CPC scheme and time-varying gain adaptive strategy, we construct an adaptive CPC controller, which guarantees the ultimate boundedness of the system, and all constraints are satisfied simultaneously during the whole control process. Finally, numerical simulation results are presented to show the effectiveness of the proposed framework.
	\end{abstract}
	
	
	\section{INTRODUCTION}
The attitude control problem of spacecraft has raised significant attention due to its engineering application background and theoretical research value. Particularly, there has been an additional emphasis on addressing the attitude control problem under motion constraints \cite{kjellberg2013discretized,kim2004constrained}, including angular velocity limitation and input saturation. 
In actual spacecraft missions, it is often necessary for spacecraft to maintain its angular velocity beneath a specified upper bound to ensure the proper functioning of attitude determination sensors, like the gyroscope and the star sensor \cite{kjellberg2013discretized}. Additionally, the control actuator equipped on the spacecraft has a maximum output capacity that cannot be exceeded. Furthermore, contemporary spacecraft missions often demand a high level of control accuracy, and thus often accompanied by additional performance requirements that must be achieved. 
Motivated by this topic, this paper investigates the spacecraft attitude control problem under multiple complex constraints, accounting for angular velocity limitation, input saturation, and fulfillment of performance requirements. Additionally, the impact of time-varying parameter uncertainty is concurrently considered in this paper.

In addressing the challenge posed by angular velocity limitation and input saturation, planning-based methodologies, such as the model predictive control \cite{kalabic2014constrained,mammarella2019attitude} and other motion planning techniques \cite{kim2004quadratically,walsh2018constrained}, employing numerical optimization techniques to generate suitable reference or feed-forward control signal sequences, thereby ensuring the satisfaction of motion constraints. However, owing to the high nonlinearity of the attitude model, the resulting nonlinear programming problem (NLP) is notoriously challenging to solve. Therefore, this greatly motivates the development of addressing motion constraints through a nonplanning method.

For the nonplanning-based methodology that is used to address the angular velocity constraint and bounded control torque, many nonlinear control techniques are utilized for this issue, as stated in \cite{hu2017unified,hu2013robust,shen2018rigid}.
In \cite{shen2018rigid}, the author delivers a barrier Lyapunov function that is constructed related to velocity states, ensuring the satisfaction of the angular velocity limitation. Nevertheless, since this method utilizes a potential-field-like method to yield constraint ability, its gradient will tend to be infinity once the maximum angular velocity is about to be reached, therefore making it hard to make full use of its allowed maximum maneuvering ability, else it may causing bad transient behavior. 
% In [REF], a time-varying sliding mode variable is defined and restricted during the whole control process. By leveraging the boundedness characteristic of quaternion, this indirectly ensures the boundedness of the angular velocity. 
 In \cite{li2016adaptive}, the author presents a method to constrain the angular velocity by using the hyperbolic tangent type function to yield an explicitly bounded virtual control law.
  However, this method actually works by enhancing the region where the virtual control law will be saturated, thus it may causing chattering at the steady state stage.
Meanwhile, several noteworthy studies in the existing literature have effectively addressed input saturation concerns, like in \cite{hu2018adaptive,li2016extended,sun2018disturbance}. 
In \cite{zou2017finite}, the author introduces an augmented system dynamics approach to "filtering" the input signal, transforming it into a bounded one that adheres to the specified saturation value. A dead-zone characteristic-based model is presented in \cite{yue2023systematic} to address the input saturation by enhancing the control effort where the system is not saturated, thereby guaranteeing the system's stability. In \cite{bang2003large}, an anti-windup auxiliary system is carried out to generate a compensation signal for input saturation. Many types of nonlinear functions, wholly or piece-wisely defined, are also developed to approximate the non-smooth saturation characteristic, as stated in \cite{boiskovic2001robust,boskovic2004robust}.

	For the performance-constrained attitude control problem, the prescribed performance control (PPC) scheme is often adopted, as presented in \cite{bechlioulis2008robust}. This scheme entails the designation of a performance function to quantify the performance requirement, further constraining the state trajectory into such a performance envelope to realize the desired performance requirements. The PPC scheme transforms the original constrained error state by leveraging an error transformation procedure to convert the original constrained system into an equivalently unconstrained one. By ensuring the convergence of the translated error or the Barrier Lyapunov function (BLF) defined in terms of the translated error, it becomes feasible to achieve convergence of the original error state while simultaneously satisfying the specified performance requirements. Given its efficacy, the PPC scheme has found application in numerous studies focusing on attitude control scenarios, as listed in \cite{wei2021overview,hu2017adaptive,hu2020model,shao2020adaptive}.
	
	As for addressing the time-varying parameter uncertainty issue, it remains open. This problem is challenging as the unknown varying parameter brings additional dynamics into the Lyapunov analysis, thereby rendering a  conventional adaptive strategy ineffective and making it hard to stabilize the system. Recently, a time-varying gain adaptive strategy is presented in \cite{gaudio2021parameter}. The author essentially provides feedback from the system's excitation condition, thereby providing a time-varying learning rate strategy (adaptive gain) regarding the specific system states. This greatly enhanced the convergence behavior of the system even in the presence of time-varying parameter uncertainty, and the system shows a better transient behavior, which may have a potential application in spacecraft control problems.
	
	The concerned multiple constraints problem is a topic of considerable research interest, which remains open to the best of the author's knowledge. Although this problem has already been discussed by combining the PPC scheme with typical angular velocity constraining methodologies, as previously investigated in \cite{golestani2022prescribed}. Nevertheless, we will elaborate on that a contradiction between the angular velocity constraint and typical PPC performance envelope constraint renders the ineffectiveness of such a combination. The contradicted constraint problem has been recently noticed and presented in \cite{mehdifar2022funnel}, where different output constraints are characterized by different funnel envelopes explicitly, and a method is presented to address those "explicitly" contradicted envelopes by considering the priority of these constraints. 
	However, for the to-be discussed motion-constrained spacecraft control problem, since such a contradiction appears during the control process and does not explicitly exhibited before the control started, thus it is hard to characterize and handle it in advance like done in \cite{mehdifar2022funnel}.
%	 Consequently, there is still lack of an efficient way to handle this issue.
	
	Based on these analysis, the main contribution of this paper is provided as follows:
	we present a framework called Compatible Performance Control (CPC) to address the spacecraft control problem under motion constraints and performance requirement. Our framework integrates the well-established Prescribed Performance Control (PPC) scheme with a strategy for detecting and mitigating contradictions. Firstly, by incorporating the concept of Zeroing Barrier Function (ZBF), we propose a detection strategy that assesses the compatibility between the angular velocity constraint and the performance envelope constraint. Subsequently, if a contradiction is identified, we introduce a projection operator with a flexible upper bound to generate a modified performance envelope signal, thereby alleviating the contradiction problem.
	Further, to overcome the limitations imposed by the angular velocity constraint, we introduce a dynamical filter technique to construct a bounded reference velocity signal, and ensures the tracking of such a reference virtual control law to further realize the desired velocity constraint.
	In addition, we address the challenge arising from time-varying parameter uncertainties by employing a time-varying gain technique, further developing an adaptive strategy that demonstrates robustness in rejecting disturbances.

	\textit{Notations}
	This paper defines the following notations for analysis. The 2-norm of any given vector or matrix is denoted by $\|\cdot\|_{2}$. Given arbitrary $\mathbb{R}^{3}$ vector $\boldsymbol{a}\in\mathbb{R}^{3}$, $\boldsymbol{a}^{\times}$ denotes the skew-symmetric matrix of the cross product manipulation. $\text{diag}(a_{i})$ denotes the element-spanned diagonal matrix, of which the diagonal line is consisted of $a_{1},a_{2},...a_{N}$ sequentially. Similarly, $\text{vec}\left(a_{i}\right)$ denotes the element-spanned column vector, of which the element is sequentially given as $a_{1},...a_{N}$. $\mathfrak{R}_{b}$ stands for the spacecraft body-fixed frame, while $\mathfrak{R}_{i}$ stands for the earth central inertia frame, $\mathfrak{R}_{d}$ stands for the target body-fixed frame. $\boldsymbol{I}_{N}$ stands for the $\mathbb{R}^{N\times N}$ identity matrix.
	
	\section{PRELIMINARIES}
We first present some useful mathematical lemmas and definitions in this section.

\begin{definition}
Defining a convex continuous differentiable evaluation function $f(\boldsymbol{L}):\mathbb{R}^{N}\to\mathbb{R}(N\ge1)$, with $\boldsymbol{L}\in\mathbb{R}^{N}$ denotes the input column vector, expressed as follows \cite{gaudio2021parameter}:
\begin{equation}\label{Lfunc}
	f(\boldsymbol{L}) \triangleq \frac{\|\boldsymbol{L}\|_{2}^{2}-L^{2}_{m}}{2\sigma L_{m}+\sigma^{2}}
\end{equation}
where $L_{m}>0$, $\sigma>0$ are design parameters. 
According to the definition of $f(\boldsymbol{L})$, it can be observed that this function $f$ is monotonically-increased as $\|\boldsymbol{L}\|_{2}$ increases such that $\lim_{\|\boldsymbol{L}\|\to+\infty}f(\boldsymbol{L})\to+\infty$ holds. Meanwhile, for $\|\boldsymbol{L}\|_{2} = L_{m}$, one has $f(\boldsymbol{L}) = 0$, while for $\|\boldsymbol{L}\|_{2} = L_{m}+\sigma$, one has $f(\boldsymbol{L}) = 1$.
\end{definition}
For the defined function $f(\boldsymbol{L}):\mathbb{R}^{N}\to\mathbb{R}$, one has the following property.
\begin{lemma}\label{LemmaPartial}
	Given arbitrary positive constant $\delta>0$.
	Let $\boldsymbol{L}_{1}\in\mathbb{R}^{N}$ be an interior point such that $f(\boldsymbol{L}_{1}) \le \delta$ holds, while $\boldsymbol{L}_{0}\in\mathbb{R}^{N}$ is an input vector satisfies $f(\boldsymbol{L}_{0}) = \delta$. Then $\boldsymbol{\nabla}^{\text{T}}_{f}(\boldsymbol{L}_{0})\left[\boldsymbol{L}_{0}-\boldsymbol{L}_{1}\right] \ge 0$ will be always hold.
\end{lemma}
The proof of Lemma \ref{LemmaPartial} is simply provided by utilizing the convexity of $f$. Since $f$ is convex, thus we have: $f(\lambda\boldsymbol{L}_{1} + (1-\lambda)\boldsymbol{L}_{0}) = f(\boldsymbol{L}_{0} + \lambda(\boldsymbol{L}_{1} - \boldsymbol{L}_{0})) \le f(\boldsymbol{L}_{0}) + \lambda(f(\boldsymbol{L}_{1})-f(\boldsymbol{L}_{0}))$ with $\lambda\in\left[0,1\right]$. Accordingly, $\left[\boldsymbol{L}_{0}-\boldsymbol{L}_{1}\right]^{\text{T}}\boldsymbol{\nabla}_{f}(\boldsymbol{L}_{0})$ can be expressed as:
\begin{equation}
	\begin{aligned}
	\left[\boldsymbol{L}_{0}-\boldsymbol{L}_{1}\right]^{\text{T}}\boldsymbol{\nabla}_{f}(\boldsymbol{L}_{0}) &= \lim_{\lambda\to0}\frac{f(\boldsymbol{L}_{0}) - f(\boldsymbol{L}_{0} + \lambda(\boldsymbol{L}_{1}-\boldsymbol{L}_{0}))}{\lambda}\\
	&\ge f(\boldsymbol{L}_{0}) - f(\boldsymbol{L}_{1}) \ge 0
	\end{aligned}
\end{equation}
Therefore, this completes the proof.

%Subsequently, based on the defined function $f$, the definition of the projection operator for the general column vector is provided as follows.
\begin{definition}\label{Proj1}
 Let $\boldsymbol{L} = \left[L_{1},...L_{N}\right]^{\text{T}}\in\mathbb{R}^{N}$, $\boldsymbol{Y}= \left[Y_{1},...Y_{N}\right]^{\text{T}}\in\mathbb{R}^{N}$ be two column vectors, then the projection operator $\text{Proj}(\cdot,\cdot,\cdot)$ for general column vector is defined as follows \cite{gaudio2021parameter}:
	\begin{equation}\label{Defproj}
	\text{Proj}(\boldsymbol{L},\boldsymbol{Y},f) \triangleq 
	\begin{cases}
		\boldsymbol{Y}- \frac{\boldsymbol{\nabla}_{f}\boldsymbol{\nabla}^{\text{T}}_{f} \cdot f(\boldsymbol{L})}{\|\boldsymbol{\nabla}_{f}\|_{2}^{2}}\boldsymbol{Y},\\
		\text{for}\quad f(\boldsymbol{L})>0 \text{and}\boldsymbol{\nabla}_{f}^{\text{T}}\boldsymbol{Y}> 0\\
	     \boldsymbol{Y},\quad\text{otherwise}
	\end{cases}
\end{equation}
where $\boldsymbol{\nabla}_{f}\in\mathbb{R}^{N}$ denotes the $\mathbb{R}^{N}$ element-spanned column vector that consisted of partial-derivatives of $f(\boldsymbol{L})$ with respect to each $i$-th component of $\boldsymbol{L}$, defined as $\boldsymbol{\nabla}f \triangleq \left[\frac{\partial f}{\partial L_{1}},...,\frac{\partial f}{\partial L_{i}},...,\frac{\partial f}{\partial L_{N}}\right]^{\text{T}}\in\mathbb{R}^{N}(i=1,2,3...N)$.
%Notably, it can be observed that $\text{Proj}(\boldsymbol{L},\boldsymbol{Y},f)$ is a $\mathbb{R}^{N}$ column vector, of which the dimension is the same as $\boldsymbol{Y}$.
\end{definition}

	\section{PROBLEM FORMULATION}
	\subsection{System Modeling}
	Considering the attitude system of a rigid-body spacecraft expressed in the unit error attitude quaternion $\boldsymbol{q}_{e}\in\mathbb{R}^{4}$, the error kinematics and dynamics model can be formulated as follows \cite{hu2017adaptive}:
	\begin{equation}\label{attsys}
		\begin{aligned}
			\dot{\boldsymbol{q}}_{ev} &= \boldsymbol{\varGamma}_{e}\boldsymbol{\omega}_{e}\\
			\dot{q}_{e0} &= -\frac{1}{2}\boldsymbol{q}^{\text{T}}_{ev}\boldsymbol{\omega}_{e}\\ 
			\boldsymbol{J}\dot{\boldsymbol{\omega}}_{e} &= \boldsymbol{J}\boldsymbol{\omega}^{\times}_e\boldsymbol{C}_e\boldsymbol{\omega}_d 
			- \boldsymbol{J}\boldsymbol{C}_e\dot{\boldsymbol{\omega}}_d
			-\boldsymbol{\omega}_s^{\times}\boldsymbol{J}\boldsymbol{\omega}_s\\
			&\quad  + \boldsymbol{\tau} + \boldsymbol{d}_{e}
		\end{aligned}
	\end{equation}
	where $\boldsymbol{q}_{ev} = \left[q_{ev1},q_{ev2},q_{ev3}\right]^{\text{T}}\in\mathbb{R}^{3}$ and $q_{e0}\in\mathbb{R}$ denotes the vector part and the scalar part of $\boldsymbol{q}_{e}$, respectively. Accordingly, $\boldsymbol{q}_{e}$ can be constructed as $\boldsymbol{q}_{e} = \left[\boldsymbol{q}^{\text{T}}_{ev},q_{e0}\right]^{\text{T}}$. 
	Subsequently, $\boldsymbol{\omega}_{e}\triangleq\boldsymbol{\omega}_{s} - \boldsymbol{C}_{e}\boldsymbol{\omega}_{d}$ stands for the error angular velocity of the spacecraft, expressed in $\mathfrak{R}_{b}$, $\boldsymbol{\omega}_{s}\in\mathbb{R}^{3}$ represents the current body-fixed angular velocity, expressed in $\mathfrak{R}_{b}$, $\boldsymbol{\omega}_{d}\in\mathbb{R}^{3}$ represents the desired angular velocity of the target body-fixed frame $\mathfrak{R}_{d}$, expressed in $\mathfrak{R}_{d}$, $\boldsymbol{C}_{e}\in\mathbb{R}^{3\times 3}$ denotes the coordinate transformation matrix from frame $\mathfrak{R}_{d}$ to $\mathfrak{R}_{b}$. $\boldsymbol{J}\in\mathbb{R}^{3\times 3}$ represents the unknown spacecraft's total inertia matrix,
	$\boldsymbol{\tau}\in\mathbb{R}^{3}$ denotes the actual exerted control input of the system, while $\boldsymbol{\varGamma}_{e} = \frac{1}{2}\left[q_{e0}\boldsymbol{I}_{3} + \boldsymbol{q}^{\times}_{ev}\right]\in\mathbb{R}^{3\times 3}$ stands for the Jacobian matrix of the kinematics equation.
	
	We then consider the parameter uncertainty.
	Let $\boldsymbol{J}_{0}\in\mathbb{R}^{3\times 3}$ denotes the known nominal inertia matrix and further defining the unknown inertia parameter uncertainty as $\Delta\boldsymbol{J}\triangleq \boldsymbol{J} - \boldsymbol{J}_{0}$, then a dynamical coupling perturbation term can be defined as $\boldsymbol{d}_{a} \triangleq - \Delta\boldsymbol{J}\dot{\boldsymbol{\omega}}_{e} - \Delta\dot{\boldsymbol{J}}\boldsymbol{\omega}_{e}$.
	We then further considering the input saturation issue. Let $\boldsymbol{u}_{c}\in\mathbb{R}^{3}$ be the calculated control input, defining the deviation between $\boldsymbol{\tau}$ and $\boldsymbol{u}_{c}$ as $\Delta\boldsymbol{u}$ such that $\Delta\boldsymbol{u} \triangleq  \boldsymbol{\tau} - \boldsymbol{u}_{c}$ holds. By sorting these notations, the dynamics equation given in equation (\ref{attsys}) can be further rearranged as follows:
	\begin{equation}\label{dyna}
		\begin{aligned}
			\boldsymbol{J}_{0}\dot{\boldsymbol{\omega}}_{e} &= \boldsymbol{\Omega}_{e} + \boldsymbol{\Omega}_{J} + \boldsymbol{u}_{c} + \Delta\boldsymbol{u} + \boldsymbol{d}
		\end{aligned}
	\end{equation}
	where $\boldsymbol{d} \triangleq \boldsymbol{d}_{a} + \boldsymbol{d}_{e}$ denotes the lumped perturbation term, $\boldsymbol{\Omega}_{e} \triangleq \boldsymbol{J}_{0}\boldsymbol{\omega}^{\times}_e\boldsymbol{C}_e\boldsymbol{\omega}_d 
	- \boldsymbol{J}_{0}\boldsymbol{C}_e\dot{\boldsymbol{\omega}}_d
	-\boldsymbol{\omega}_s^{\times}\boldsymbol{J}_{0}\boldsymbol{\omega}_s\in\mathbb{R}^{3}$ denotes the dynamical-coupling term that can be directly calculated, while $\boldsymbol{\Omega}_{J}\triangleq \Delta\boldsymbol{J}\boldsymbol{\omega}^{\times}_{e}\boldsymbol{C}_{e}\boldsymbol{\omega}_{d} - \Delta\boldsymbol{J}\boldsymbol{C}_{e}\dot{\boldsymbol{\omega}}_{d} - \boldsymbol{\omega}^{\times}_{s}\Delta\boldsymbol{J}\boldsymbol{\omega}_{s}\in\mathbb{R}^{3}$ represents the unknown dynamical-coupling term.
	
	Further, we employ the linear operator $\mathcal{L}(\cdot): \mathbb{R}^{3}\to\mathbb{R}^{3\times 6}$ that utilized in \cite{hu2018adaptive} to transform the dynamical system (\ref{dyna}) into a standard linear-parameterized plant. Defining the regression matrix $\boldsymbol{W}_{J}\in\mathbb{R}^{3\times 6}$ and the estimate vector $\boldsymbol{\Theta}_{J} \in\mathbb{R}^{6}$ as follows:
	\begin{equation}
		\begin{aligned}
			\boldsymbol{W}_{J} &\triangleq \mathcal{L}(\boldsymbol{\omega}^{\times}_{e}\boldsymbol{C}_{e}\boldsymbol{\omega}_{d} - \boldsymbol{C}_{e}\dot{\boldsymbol{\omega}}_{d}) - \boldsymbol{\omega}^{\times}_{s}\mathcal{L}(\boldsymbol{\omega}_{s})\\
			\boldsymbol{\Theta}_{J} &\triangleq \left[\Delta_{J11},\Delta_{J12},\Delta_{J13},\Delta_{J22},\Delta_{J23},\Delta_{J33}\right]^{\text{T}} 
		\end{aligned}
	\end{equation}
	Subsequently, defining a dimensional-expanded regression matrix and an estimation vector as $\boldsymbol{W}\in\mathbb{R}^{3\times 9}$ and $\boldsymbol{\Theta}\in\mathbb{R}^{9\times 1}$, expressed as follows:
	\begin{equation}\label{Wexpand}
		\begin{aligned}
			\boldsymbol{W} \triangleq \left[\boldsymbol{W}_{J},\boldsymbol{I}_{3}\right]\in\mathbb{R}^{3\times 9},
			\boldsymbol{\Theta} \triangleq \left[\boldsymbol{\Theta}^{\text{T}}_{J},\boldsymbol{d}^{\text{T}}\right]^{\text{T}}\in\mathbb{R}^{9}
		\end{aligned}
	\end{equation}
	Accordingly, applying the defined $\boldsymbol{W}$ and $\boldsymbol{\Theta}$, equation (\ref{dyna}) can be further integrated as:
	\begin{equation}
		\boldsymbol{J}_{0}\dot{\boldsymbol{\omega}}_{e} = \boldsymbol{W}\boldsymbol{\Theta} +  \boldsymbol{\Omega}_{e}+\boldsymbol{u}_{c} + \Delta\boldsymbol{u}	
	\end{equation}

	For the synthesize of the following designing process, we introduce the following assumptions and definitions:
	
	\begin{assumption}\label{KnownJ0}
		(Known Nominal Matrix) Assuming that the nominal inertia matrix $\boldsymbol{J}_{0}\in\mathbb{R}^{3\times 3}$ is known, symmetric and positive-definite. Accordingly, let $\lambda_{J0\max}$, $\lambda_{J0\min}$ be the maximum and minimum eigenvalues of $\boldsymbol{J}_{0}$, one has $\lambda_{J0\min}\boldsymbol{x}^{\text{T}}\boldsymbol{x} \le \boldsymbol{x}^{\text{T}}\boldsymbol{J}_{0}\boldsymbol{x} \le \lambda_{J0\max}\boldsymbol{x}^{\text{T}}\boldsymbol{x}$ for $\forall \boldsymbol{x}\in\mathbb{R}^{3}$.
	\end{assumption}
	\begin{assumption}\label{unknownJ}
	     Assuming that the total unknown inertia matrix $\boldsymbol{J}\in\mathbb{R}^{3\times 3}$ and the unknown inertia matrix discrepancy  $\Delta\boldsymbol{J}\in\mathbb{R}^{3\times 3}$ are both symmetric positive-definite matrices.
	\end{assumption}
		\begin{assumption}\label{BoundedUnknown}
		(Bounded Parameter Uncertainty and its Time-Varying Rate)
		Assuming that the unknown parameter $\|\boldsymbol{\Theta}\|_{2}$ along with its time-derivative $\|\dot{\boldsymbol{\Theta}}\|_{2}$ are bounded by positive constants, such that $\|\boldsymbol{\Theta}\|\le \Theta_{\max}$ and $\dot{\boldsymbol{\Theta}}\le \Theta_{d\max}$ holds for $\forall t\in\left[t_{0},+\infty\right)$.
	\end{assumption}
	\begin{assumption}\label{DisAss}
		(Bounded Disturbance) For the lumped perturbation term $\boldsymbol{d}$, assuming that exists a positive constant $D_{m}$ such that $\|\boldsymbol{d}\| \le D_{m}$ holds for $\forall t\in\left[t_{0},+\infty\right)$.
	\end{assumption}
	\begin{remark}
		Since the angular velocity constraint is considered in this paper, therefore there exists upper boundary for $\boldsymbol{\omega}_{e}$, $\boldsymbol{\omega}_{d}$ along with its time-derivatives. Meanwhile, owing to the boundedness of parameter uncertainty, as presented by Assumption \ref{BoundedUnknown}, this further indicates that a sufficiently large upper bound $D_{m}$ exists for $\boldsymbol{d}$, and such Assumption \ref{DisAss} is reasonable.
	\end{remark}

	\subsection{Constraint Description}\label{CONSDES}
	In this paper, we mainly consider three types of constraints: the angular velocity limitation, the input saturation and the performance envelope constraint.
	\subsubsection{Angular Velocity Limitation}
	During the whole control process, each $i$-th component of the spacecraft's angular velocity $\boldsymbol{\omega}_{s}$ should remain beneath a given constant $\Omega_{\max}$, i.e., the following inequality holds:
	\begin{equation}\label{angcons}
		|\omega_{si}(t)| \le \Omega_{\max},\quad \text{for} \quad t\in\left[t_{0},+\infty\right)(i = 1,2,3)
	\end{equation} 
	where $\omega_{si}(t)$ represents the $i$-th component of $\boldsymbol{\omega}_{s}$, $\Omega_{\max} > 0$ stands for the given upper bound.
	\subsubsection{Input Saturation Constraint}
	The relationship between the actual control input signal $\boldsymbol{\tau}$ and the command 
	control input $\boldsymbol{u}_{c}$ satisfies the following saturation characteristic, expressed as follows:
	\begin{equation}\label{inputcons}
		\tau_{i} = 
		\begin{cases}
			u_{ci},&\quad |u_{ci}|\le U_{\max}\\
			U_{\max}\text{sgn}\left(u_{ci}\right),&\quad |u_{ci}| > U_{\max}
		\end{cases}
	\end{equation}
	where $\tau_{i}$, $u_{ci}$ denotes the $i$-th component of $\boldsymbol{\tau}$ and $\boldsymbol{u}_{c}$, respectively.
	
	\subsubsection{Performance Envelope Constraint}
	
	Following the fundamental idea in Prescribed Performance Control (PPC), in order to satisfy the desired performance criteria, the time-evolution of $q_{evi}(t)(i=1,2,3)$ should be restricted in a sector-like region that is enclosed by a pair of performance functions, denoted as $\rho_{i}(t)$ and $-\rho_{i}(t)$, with $\rho_{i}(t) > 0$ holds. Therefore, the performance envelope constraint can be established as follows:
	\begin{equation}\label{percons}
		q_{evi}(t) \in\left(-\rho_{i}(t) , \rho_{i}(t)\right),\quad\text{for}\quad t\in\left[t_{0},+\infty\right)(i=1,2,3)
	\end{equation}
	where $q_{evi}(t)$ represents the $i$-th component of $\boldsymbol{q}_{ev}$, $\rho_{i}(t)$ represents the performance envelope that corresponds to the $i$-th error state.
	
	\subsection{Control Objective}\label{ControlObj}
	The main control objective of this paper is to develop a controller that guides the attitude error system given in equation (\ref{attsys}) to its equilibrium point $\boldsymbol{q}_{ev} = \boldsymbol{0}$, $\boldsymbol{\omega}_{e} = \boldsymbol{0}$ in the presence of time-varying parameter uncertainties, while all these constraints given in equation (\ref{angcons})(\ref{inputcons}) and (\ref{percons}) are satisfied during the whole control process.

	\section{COMPATIBLE PERFORMANCE CONTROL (CPC) SCHEME DESIGN}\label{MAINRESUlT}
	%In order to address the presented control objective given in Subsection \ref{ControlObj}, 
	%
	%Firstly, A dynamical filter is introduced to address the angular velocity limitation, as elaborated in Subsection \ref{SVG}. This filter generates a virtual control reference signal that satisfies the given upper bound. By ensuring that the actual state tracks this reference signal, the system will converges with the satisfaction of the angular velocity limitation. 
	%
	%
	%Subsequently, in Subsection \ref{PFdesign}.\ref{detection}, the concept of Zeroing Barrier Function (ZBF) is employed to construct a contradiction detection, and an indicator function is further designed to judge whether the current performance envelope is compatible with the angular velocity limitation or not. Additionally, a feedback auxiliary system is designed in Subsection \ref{PFdesign}.\ref{MODPPF} to generate a modification signal, expanding the constraint region when such a contradiction exists.

	\subsection{Motivation of Compatible Performance Control}\label{Motivation}
	As discovered in existing literature, the Prescribed Performance Control (PPC) scheme has been widely used to achieve specific performance requirements. This scheme involves designing a performance function and constraining the state trajectory to remain within the performance envelope, in order to satisfy the desired performance requirements. Additionally, there exist numerous effective frameworks for attitude control under angular velocity constraint.
	For the presented control objective given in Subsection \ref{ControlObj}, it seems that a trivial combination between the PPC scheme and other angular velocity constraining methodologies provide a promising solution \cite{golestani2022prescribed}. Nevertheless, this issue is not as behoove as it seems. We will elaborate on that a potential contradiction between the angular velocity limitation and the performance envelope constraint will render the ineffectiveness of such a direct combination.
	
\begin{figure}[hbt!]
		\centering 
		\includegraphics[width=0.5\textwidth]{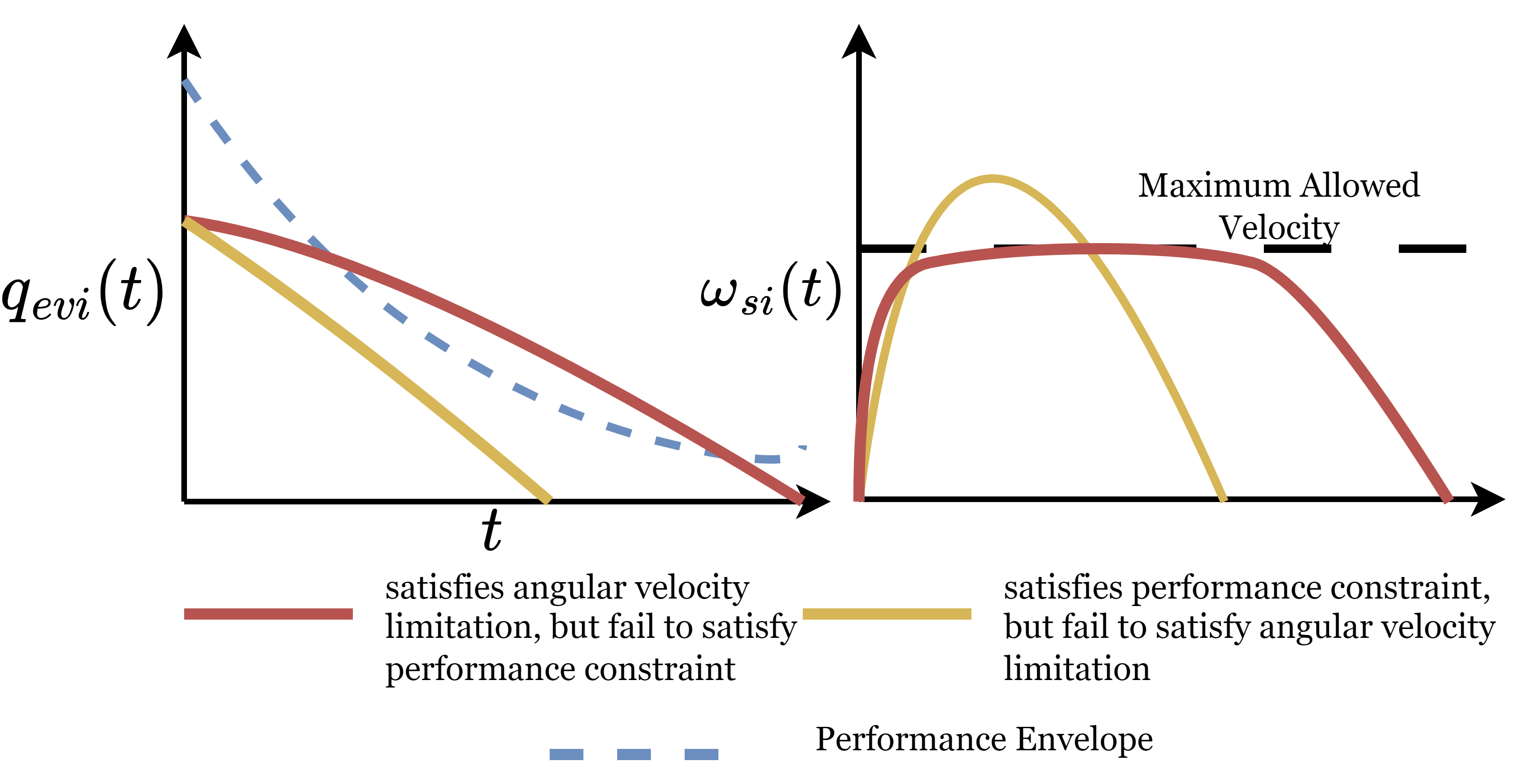}
		\caption{Sketch map of the intrinsic conflict between the angular velocity limitation and performance  envelope constraint}       
		\label{conflict}   
	\end{figure}

	As illustrated in Figure \ref{conflict}, for any given $i$-th component of the error state $\boldsymbol{q}_{ev}(t)$, there exists a maximum angular velocity limitation that must be respected during arbitrary time interval $t\in[t_{0},t_{1}](t_{0}<t_{1})$. As a result, it can be inferred that the possible state trajectory that converges the fastest for $q_{evi}(t)$ is the one that corresponds to the circumstance where spacecraft rotates at its maximum allowable angular velocity, referring to the red-solid line in Figure \ref{conflict}.
	However, if the performance envelope (the blue dotted curve in Figure \ref{conflict}), which is generally designed without consideration of motion constraints, is designed to converge too quickly, then the resulting state trajectory may fall outside of the performance envelope, even if it converges at the fastest possible angular velocity. One can be discovered that due to this incompatibility problem, the system may impossible to satisfy both two constraints simultaneously under some circumstances. On the other hand, although it may possible to assign a performance function with slow convergence behavior, this may be too conservative and thus may causing a degrading on system's performance.
	
		Given the importance of adhering to angular velocity limitations in actual spacecraft missions, it is crucial to prioritize such safety concerns over a preassigned performance envelope constraint. As such, the angular velocity limitation should be guaranteed to be met, even if it requires a concession of the preassigned performance envelope constraint.
	Meanwhile, once the issue of conflicting constraints has been resolved, the performance envelope constraint should recover to the original design, and the system state should be guided to meet the originally desired performance criteria. 
	
\begin{figure}[hbt!]
		\centering 
		\includegraphics[width=0.5\textwidth]{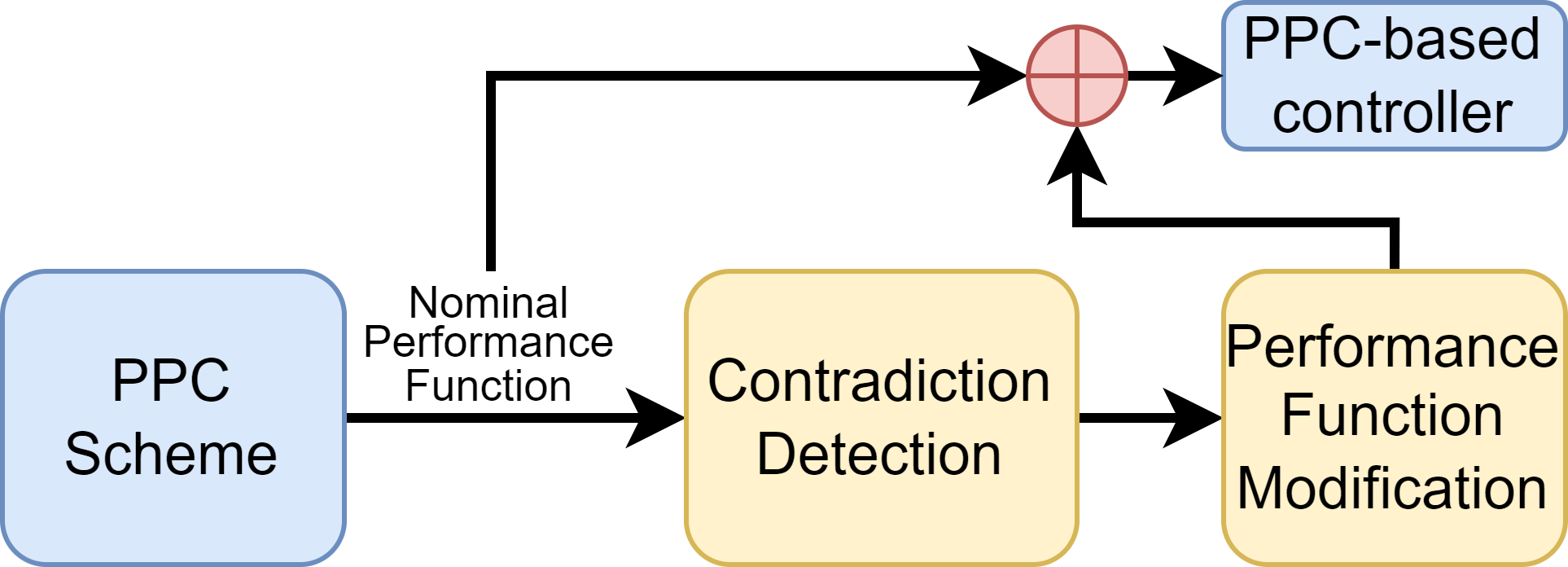}
		\caption{Schematic Diagram of the proposed Compatible Performance Control Scheme}       
		\label{CPC}   
	\end{figure}

		Following this idea, this paper proposes a Compatible Performance Control (CPC) scheme, which integrates the conventional PPC framework with a contradiction detection and alleviation strategy. Firstly, we employ the concept of Zeroing Barrier Function (ZBF) to construct a contradiction detection strategy, showing whether the preassigned performance envelope constraint is compatible with the angular velocity limitation or not, and thus determining whether the aforementioned "concession" on preassigned performance envelope constraint is necessary or not, detailed in Subsection \ref{detection}. Then, a dynamic projection-operator-based dynamical system is further carried out to provide a modification signal, rapidly expanding the performance envelope temporarily and thus alleviating the potential contradiction, detailed in Subsection \ref{MODPPF}.
	A structural scheme diagram of the CPC scheme is illustrated in Figure \ref{CPC}.

%	The CPC scheme will prioritize the angular velocity limitation, which is a safety concern, over the performance envelope constraint when such a contradiction exists, and recover the performance envelope to the original design when such a contradiction vanishes.
	
	\subsection{System Transformation}\label{systrans}
	Regarding the typical Prescribed Performance Control (PPC) scheme, for arbitrary $i$-th component of $\boldsymbol{q}_{ev}(t)$, the following constraint should be satisfied:
	\begin{equation}\label{errcons}
		-\rho_{i}(t) < q_{evi}(t) < \rho_{i}(t)(i=1,2,3)
	\end{equation}
	where $\rho_{i}(t)>0$ denotes the corresponding performance function for $q_{evi}(t)$, which will be designed later in Subsection \ref{PFdesign}.
	Defining a unified error variable as:
	\begin{equation}
		\varepsilon_{i}(t) \triangleq \frac{q_{evi}(t)}{\rho_{i}(t)}
	\end{equation}
then the error constraint in equation (\ref{errcons}) can be equivalently rearranged as $-1<\varepsilon_{i}(t)<1(i = 1,2,3)$.
	Therefore, taking the time-derivative of the translated error variable that defined as $\boldsymbol{\varepsilon}(t) = \left[\varepsilon_{1}(t),\varepsilon_{2}(t),\varepsilon_{3}(t)\right]^{\text{T}}\in\mathbb{R}^{3}$, the error kinematics given by equation (\ref{attsys}) can be transformed as follows:
	\begin{equation}\label{outputsys}
		\dot{\boldsymbol{\varepsilon}} = \boldsymbol{\Psi}\left[\dot{\boldsymbol{q}}_{ev} - \boldsymbol{\Lambda}\boldsymbol{q}_{ev}\right]
	\end{equation}
	where $\boldsymbol{\Psi} \triangleq \text{diag}(1/\rho_{i}(t))\in\mathbb{R}^{3\times 3}$, $\boldsymbol{\Lambda}\triangleq\text{diag}\left(\dot{\rho}_{i}(t)/\rho_{i}(t)\right)\in\mathbb{R}^{3\times 3}$ are two defined diagonal matrices. Notably, recalling the definition of $\boldsymbol{\varepsilon}$, we have $\boldsymbol{\Psi}\boldsymbol{q}_{ev} = \boldsymbol{\varepsilon}$.
	
	In order to facilitate the following analysis, by following the backstepping control methodology, we further let $\boldsymbol{\omega}_{v}\in\mathbb{R}^{3}$ be the virtual angular velocity that stabilizes the transformed output system (\ref{outputsys}). Correspondingly, the total system is converted to the following two cascaded error subsystem, defined as:
	\begin{equation}\label{errorsub}
		\boldsymbol{z}_{1} = \boldsymbol{\varepsilon},\boldsymbol{z}_{2} = \boldsymbol{\omega}_{e} - \boldsymbol{\omega}_{v}
	\end{equation}

	\subsection{Bounded Velocity Generator}\label{SVG}
	Motivated by \cite{li2016adaptive}, this paper delivers a layered strategy to satisfy the angular velocity limitation: we first design each $i$-th component of the virtual control law $\boldsymbol{\omega}_{v}$ that stabilizes the transformed output system $\boldsymbol{\varepsilon}$ to be explicitly bounded. Then, by ensuring the convergence of backstepping tracking error $\boldsymbol{z}_{2}$, the deviation between actual $\boldsymbol{\omega}_{e}$ and $\boldsymbol{\omega}_{v}$ vanishes, and thus ensuring the bounded characteristic of $\boldsymbol{\omega}_{s}$.
	
	Inspired by the methodology utilized for addressing input saturation issues as referenced in \cite{zou2017finite}, we adopt this dynamical-filter technique to serve as a bounded velocity generator, generating bounded $\boldsymbol{\omega}_{v}$ from arbitrary given stabilizing law.
	For arbitrary $i$-th component, the filter system is given as follows:
	\begin{equation}\label{filter}
		\dot{\omega}_{vi}(t) = -C_{d}C_{f}\omega_{vi}(t) + C_{f}\mathcal{K}(\omega_{vi}(t),B_{\omega})v_{i}(t)
	\end{equation}
	where $\omega_{vi}(t)$, $v_{i}(t)$ represents the $i$-th component of the filter output $\boldsymbol{\omega}_{v}\in\mathbb{R}^{3}$ and the input $\boldsymbol{v}\in\mathbb{R}^{3}$, correspondingly,
	$C_{f}, C_{d}>0$ are positive constant coefficients, $B_{\omega}>0$ is a given parameter that selected considering the given angular velocity limitation $\Omega_{\max}$, $\mathcal{K}(\omega_{vi}(t),B_{\omega})$ is a function with following properties:
	\begin{equation}
		\begin{aligned}
			\lim_{|\omega_{vi}(t)|\to B_{\omega}}\mathcal{K}(\omega_{vi}(t),B_{\omega})&=0,\quad \mathcal{K}(0,B_{\omega}) = 1\\
			\frac{\partial \mathcal{K}(\omega_{vi},B_{\omega})}{\omega_{vi}} < 0&, \text{for}\quad\omega_{vi}(t)\in\left[0,B_{\omega}\right)
		\end{aligned}
	\end{equation}

		Notably, owing to the fact that $\lim_{|\omega_{vi}(t)|\to B_{\omega}}\dot{\omega}_{vi}(t)=-C_{d}C_{f}\omega_{vi}(t) < 0$ holds, given the initial condition as $\omega_{vi}(t_{0}) = 0$ and assumes that $|v_{i}(t)|$ will not tend to be $+\infty$, by utilizing the filter in equation (\ref{filter}), it yields a bounded output signal $\boldsymbol{\omega}_{v}$ such that $|\omega_{vi}(t)| < B_{\omega}$ holds for $\forall t\in\left[t_{0},+\infty\right)$, this characteristic is also stated in \cite{zou2017finite}.
%		 Therefore, let the input signal $v_{i}(t)$ be a stabilization control law without considering the upper bound, then the dynamical filter system serves as a "shaper", which shapes the input $v_{i}(t)$ to $\omega_{vi}(t)$ that strictly below $B_{\omega}$.

	\begin{remark}\label{P1}
		When the system is away from the saturation, i.e. $|\omega_{vi}(t)| \ll B_{\omega}$, it can be observed that the filter system given in equation (\ref{filter}) can be approximated as:
			$\dot{\omega}_{vi}(t) = -C_{d}C_{f}\left[\omega_{vi}(t) - \frac{1}{C_{d}}v_{i}(t)\right]$, which can be regarded as a low-pass filter for the input $\frac{1}{C_{d}}v_{i}(t)$. This indicates that the deviation between $\omega_{vi}(t)$ and $\frac{1}{C_{d}}v_{i}(t)$ will converge to a small region.
	\end{remark}
	\begin{remark}\label{Bw}
		Owing to the existence of $\boldsymbol{z}_{2}$ and the nonzero $\boldsymbol{\omega}_{d}$, the choosing of $B_{\omega}$ should set some additional margin, and $B_{\omega} < \Omega_{\max} - \max\left(|\omega_{di}(t)|\right)(i=1,2,3)$ is recommended.
	\end{remark}
	
%	\begin{remark}
%		For the general attitude tracking task, note that $\boldsymbol{\omega}_{d} \neq \boldsymbol{0}$ holds. Since $\boldsymbol{\omega}_{s} = \boldsymbol{\omega}_{e} + \boldsymbol{C}_{e}\boldsymbol{\omega}_{d}$ is satisfied all along, thus it can be yielded that $\omega_{si} = \omega_{ei} + \boldsymbol{C}_{e}(i,:)\boldsymbol{\omega}_{d}$ holds, where $\boldsymbol{C}_{e}(i,:)$ denotes the $i$-th row of $\boldsymbol{C}_{e}$. This indicates the $B_{\omega} < \Omega_{\max}$ should be set appropriately to set additional margin for non-zero $\boldsymbol{\omega}_{d}$. For the special case that $\boldsymbol{\omega}_{d} = \boldsymbol{0}$, $B_{\omega} = \Omega_{\max}$ can be directly selected.
%	\end{remark}

	\subsection{Performance Function Design, Contradiction Detection and Alleviation}\label{PFdesign}
	For each $i$-th component of $\boldsymbol{q}_{ev}(t)$, the performance function $\rho_{i}(t)$ is consists of two parts: the nominal part and the modification part, denoted as $\rho_{ni}(t)$ and $\Delta{\rho}_{i}(t)$, respectively. 
%	$\rho_{ni}(t)$ will provide a nominal converging performance envelope for the system, which is designed without considering the angular velocity limitation. $\Delta\rho_{i}(t)$ will provide a suitably large positive signal, expanding the nominal performance envelope $\rho_{ni}(t)$ when the contradiction between the angular velocity limitation and the nominal performance envelope is detected, as previously mentioned in Subsection \ref{Motivation}. Consequently, 
	Therefore, the total performance function $\rho_{i}(t)$ can be expressed as:
	\begin{equation}
		\rho_{i}(t)  = \rho_{ni}(t) + \Delta\rho_{i}(t)
	\end{equation}
	
	\subsubsection{Nominal Performance Function Design}
	The nominal performance function $\rho_{ni}(t)$ is designed as a widely-utilized exponentially-converged function, expressed as follows \cite{bechlioulis2008robust}:
	\begin{equation}\label{NPF}
		\rho_{ni}(t) = (\rho_{0i} - \rho_{\infty i})e^{-\gamma_{i} t}+\rho_{\infty i} 
	\end{equation}
	where $\rho_{0i} > 0$, $\rho_{\infty i}>0$ stands for the initial value and the asymptote value of $\rho_{ni}(t)$, respectively; $\gamma_{i} > 0$ represents the exponential index, corresponding to the $i$-th component. 
%	It can be observed that the nominal performance function is an exponentially converged one that starts from $\rho_{ni}(0) = \rho_{0i}$, and tend to be $\rho_{\infty i}$ for $t\to+\infty$. Practically speaking, $\rho_{0i}$ can be chosen according to the initial condition of $q_{evi}(t)$ as $\rho_{0i} > |q_{evi}(0)|$, which ensures that the state trajectory will be enclosed by the performance envelope at the initial stage.

	\subsubsection{Contradiction Detection Indicator}\label{detection}
	
	In the following context, we focus on the design of the modification part $\Delta\rho_{i}(t)$, further highlighting \textbf{when} the \textbf{how} it will be generated. 
	
	To determine when to generate such a modification signal, we first design a \textit{Contradiction Detection Strategy} by employing the concept of Zeroing Barrier Function (ZBF).
	Defining a judgment function as follows:
	\begin{equation}
		\mathcal{D}_{i}(t) \triangleq \rho_{ni}^2(t) - q_{evi}^2(t)
	\end{equation}
	The 0-level set of the function $\mathcal{D}_{i}(t)$ can be observed to separate the region where the \textbf{nominal} performance envelope constraint $\rho_{ni}(t)<q_{evi}(t)<\rho_{ni}(t)$ is satisfied or unsatisfied. Therefore, the function $\mathcal{D}_{i}(t)$ serves as a ZBF for the nominal performance envelope constraint. According to the theorem referenced in \cite{ames2019control}, if the following condition holds:
	\begin{equation}
		\dot{\mathcal{D}}_{i}(t) \ge -\alpha(\mathcal{D}_{i}(t))
	\end{equation}
	then the nominal performance constraint-satisfied region remains invariant over time. Here, $\alpha(\cdot)$ belongs to an extended class of $\mathcal{K}$ functions, and a positive constant $\alpha>0$ can be chosen practically. 

\begin{figure}[hbt!]
			\centering 
			\includegraphics[width=0.25\textwidth]{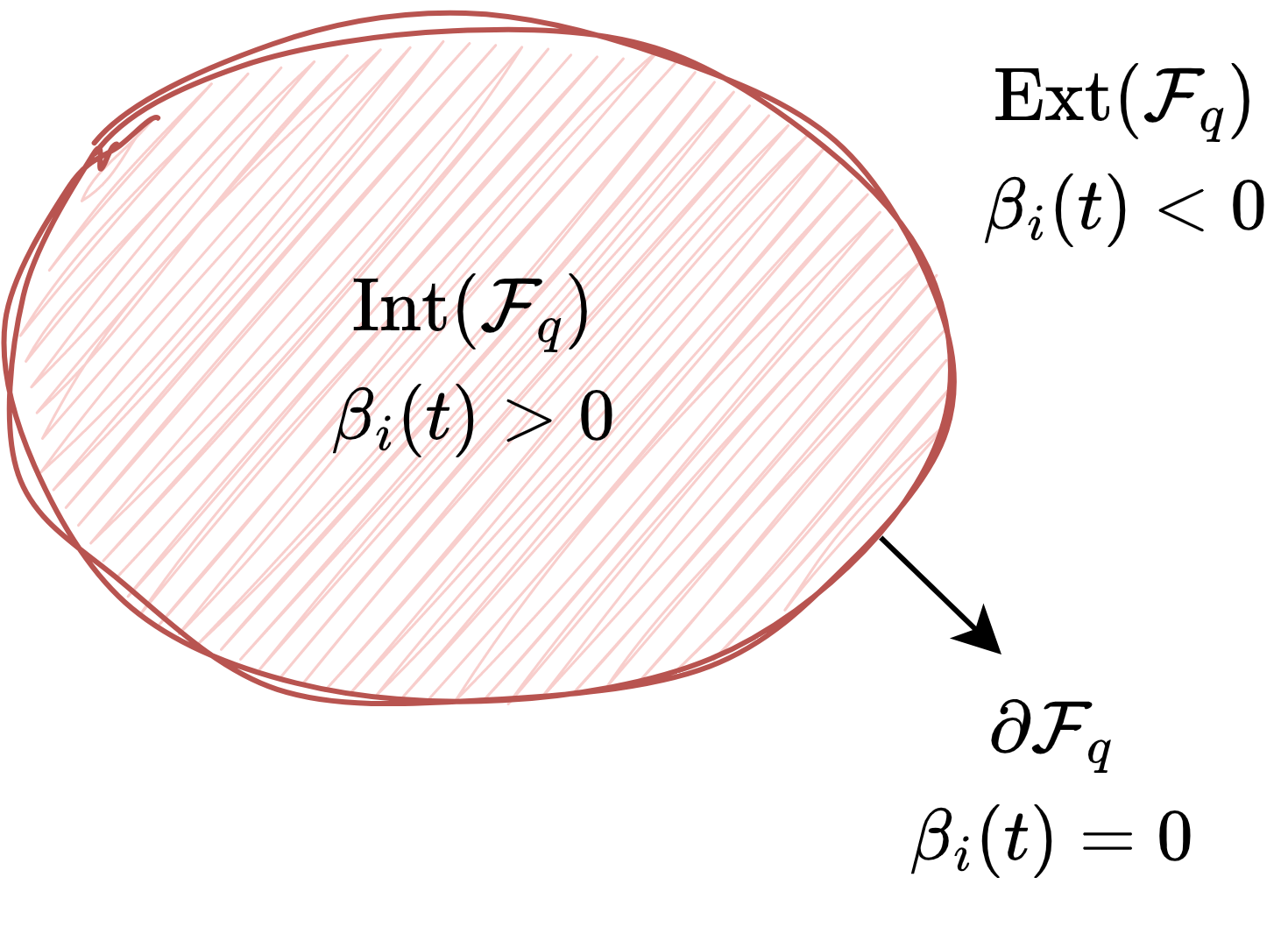}
			\caption{Relationship between the nominal performance-constraint-satisfied set $\mathcal{F}_{q}$ and the variable $\beta_{i}(t)$}       
			\label{judgment}   
		\centering 
		\includegraphics[width=0.45\textwidth]{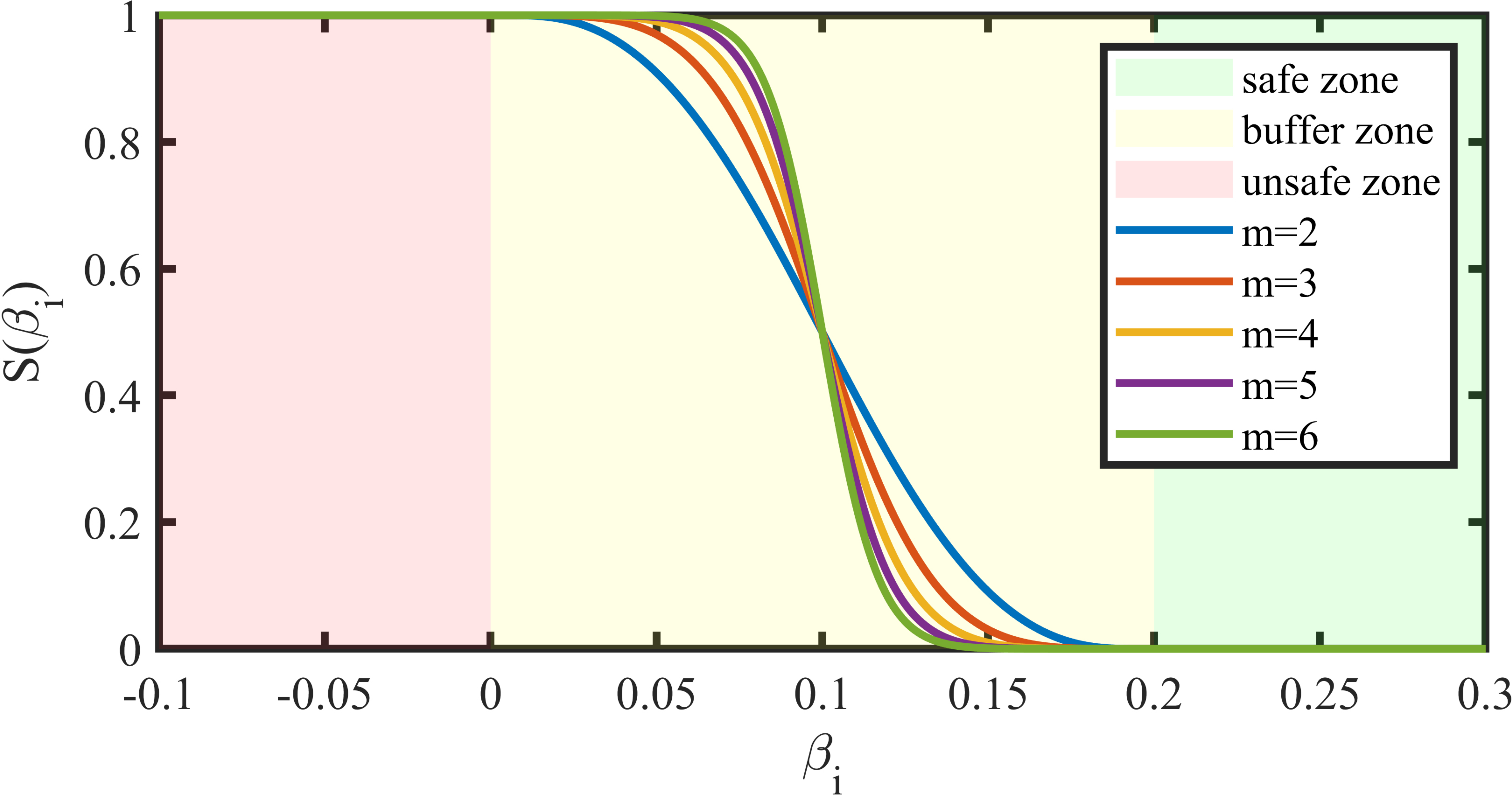}
		\caption{Example Illustration of the Indicator Function $\mathcal{S}(\beta_{i}(t))$}       
		\label{indicator}   
	\end{figure}

	Subsequently, we define a new variable $\beta_{i}(t)$ as follows $\beta_{i} \triangleq \dot{\mathcal{D}}_{i}(t) + \alpha\mathcal{D}_{i}(t)$.
	Therefore, it provides a simple judgment condition: denote the nominal performance-envelope-satisfied set as $\mathcal{F}_{q}$, then $\beta_{i}(t) > 0$ corresponds to $\text{Int}\left(\mathcal{F}_{q}\right)$, $\beta_{i}(t) = 0$ corresponds to $\partial\mathcal{F}_{q}$ and $\beta_{i}(t) < 0$ corresponds to $\text{Ext}(\mathcal{F}_{q})$, as illustrated in Figure \ref{judgment}. 
	
	Furthermore, we utilize a smooth mollified switching function to construct an $0-1$ indicator function  $\mathcal{S}(\beta_{i})$, which serves as a "Bool" variable, expressed as follows:
	\begin{equation}
		\begin{aligned}
			\label{Omegas}
			&\mathcal{S}(\beta_{i})= \begin{cases}
				1 &  \beta_{i}\in\left(-\infty,S_{0}\right)\\
				-\frac{1}{2}\left[\tanh\frac{m\left(S_{1}-S_{0}\right)\cdot\left(\beta_{i}-S_{m}\right)}{\sqrt{(\beta_{i}-S_{0})(S_{1}-\beta_{i})}}-1\right] & \beta_{i}\in\left[S_{0},S_{1}\right)\\
				0 &  \beta_{i}\in\left[S_{1},+\infty \right)\\
			\end{cases}
		\end{aligned}
	\end{equation}  
	where $S_0$, $S_1$, and $S_m$ determines when will the indicator function $\mathcal{S}(\beta_{i})$ switches; $m \in \left[\frac{1}{S_{1}-S_{0}},+\infty\right)$ controls the decreasing rate of the indicator function $\mathcal{S}(\beta_{i})$ when $\beta_{i}$ increases. To facilitate the detection, we set $S_{0} = 0$, $S_{m} = S_{0}+\epsilon$ and $S_{1} = S_{0} + 2\epsilon$, with $\epsilon > 0$ is a small constant.
	
	According to the above design, for $\beta_{i}(t) = \dot{\mathcal{D}}_{i} +\alpha\mathcal{D}_{i}(t) < S_{0}$, $\mathcal{S}(\beta_{i}(t)) = 1$ holds, this indicates that the system is suffering from the risk of contradicted constraints, and thus a modification is necessary. For $\beta_{i}(t) > S_{1}$, we have $\mathcal{S}(\beta_{i}(t)) = 0$, indicating that the nominal performance envelope is compatible with the given angular velocity limitation. For $\beta_{i}(t)\in\left[S_{0},S_{1}\right)$, it is a buffer zone, where $\mathcal{S}(\beta_{i}(t))$ varying from $1$ to $0$. For instance, given the parameter $m = 2,3,4,5,6$, $S_{0} = 0$, $\epsilon = 0.1$, an example of the indicator function $\mathcal{S}(\beta_{i}(t))$ is illustrated in Figure \ref{indicator}. 
	%The green region represents the safe zone where $\mathcal{S}(\beta_{i}(t)) = 0$, yellow region represents the buffer zone where $\mathcal{S}(\beta_{i}(t))\in\left(0,1\right)$, while red region represents the unsafe zone where $\mathcal{S}(\beta_{i}(t)) = 1$.
%	\begin{remark}
%		It can be observed that the constant $\epsilon > 0$ serves as a robust boundary, which expands the performance-constraint-unsatisfied region, thereby providing additional robustness.
%	\end{remark}
%	\begin{remark}
%		For the selection of $\alpha$, it can be observed that a larger $\alpha$ will render a smaller performance-constraint-unsatisfied region. Therefore, $\alpha$ should be set to be appropriately small to obtain an in-advance detection of the potential violation of the performance constraint. Specially, for $\alpha = 0$, it prohibits any approaching of $q^{2}_{evi}(t)$-trajectory towards the $\rho^{2}_{ni}(t)$-trajectory, however, this may be too conservative.
%	\end{remark}

	\subsubsection{Time-Varying Projection Operator-based Performance Envelope Modification }\label{MODPPF}
	%Based on the designed indicator function, it can be observed that for $\mathcal{S}(\beta_{i}(t)) = 1$, it indicates that it is necessary to generate a sufficiently large positive modification signal. 
	To facilitate the following analysis, we first defining a deviation $\boldsymbol{\varpi}(t)$ as follows:
	\begin{equation}
		\boldsymbol{\varpi} \triangleq \boldsymbol{\omega}_{v} - \frac{1}{C_{d}}\boldsymbol{v}
	\end{equation}
	where $\boldsymbol{\omega}_{v}$ and $\boldsymbol{v}$ stands for the output and the input of the bounded velocity generator system that previously designed in Subsection \ref{SVG}.
	Applying the defined $\boldsymbol{\varpi}$ and the designed indicator function $\mathcal{S}(\beta_{i}(t))$, a projection operator-governed dynamical system is designed to generate a bounded modification signal, expressed as: $\Delta\dot{\rho}_{i}(t) = \text{Proj}\left(\Delta\rho_{i}(t),\mathcal{M}(t),f(\Delta\rho_{i}(t),\rho^{i}_{\max}(t))\right)$.
    Specifically, $\Delta\dot{\rho}_{i}(t)$ is specified as follows:
	\begin{equation}\label{dotrho}
	\Delta\dot{\rho}_{i}(t) \triangleq 
	\begin{cases}
		\mathcal{M}(t)- 
		\frac{(\nabla^{i}_{\rho})^{2}f(\Delta\rho_{i}(t),\rho^{i}_{\max}(t))\mathcal{M}(t)}{\left(\nabla^{i}_{\rho}\right)^{2}}-\frac{\nabla^{i}_{\rho\max}\rho_{\max}^{i}(t)}{\nabla^{i}_{\rho}}\\
		\text{for}\quad f>0\text{and}\nabla^{i}_{\rho}\mathcal{M}(t)+\nabla^{i}_{\rho\max}\dot{\rho}^{i}_{\max}(t)>0\\
		\mathcal{M}(t)-\frac{\nabla^{i}_{\rho\max}\rho^{i}_{\max}(t)}{\nabla^{i}_{\rho}},\quad \text{otherwise}
	\end{cases}
\end{equation}
	\begin{equation}\label{modified}
	\mathcal{M}(t) = -C_{\rho}\Delta\rho_{i}(t) + C_{v}\mathcal{S}(\beta_{i}(t))|\tanh(F_{v}\varpi_{i}(t))|
	\end{equation}
		Notably, equation (\ref{dotrho}) can be further rewritten in a compact form as:
		\begin{equation}\label{dotrhocompact}
		\Delta\dot{\rho}_{i}(t) = g(t)\mathcal{M}(t)-\mathcal{W}(t)
		\end{equation}
		where $g(t)\in\left[0,1\right]$ stands for an efficiency factor, $\mathcal{W}(t)$ is an additional term, specified as:$\mathcal{W}(t)\triangleq\frac{\nabla^{i}_{\rho\max}}{\nabla^{i}_{\rho}}\rho_{\max}^{i}(t)$.
		
	In the above equation, $f(\Delta\rho_{i}(t),\rho^{i}_{\max}(t))$ denotes the evaluation function of the projection operator, which has a similar form as we defined in Definition \ref{Lfunc}, expressed as follows:
	\begin{equation}
		f(\Delta\rho_{i}(t),\rho^{i}_{\max}(t)) 
		= \frac{\Delta\rho^{2}_{i}(t)-(\rho^{i}_{\max}(t))^{2}}{2\sigma\rho^{i}_{\max}(t)+\sigma^{2}}
	\end{equation}
	where $\rho^{i}_{\max}(t)$ is a time-varying upper bound of the projection operator, given as:
	\begin{equation}\label{rhomax}
		\rho^{i}_{\max}(t) =
		\begin{cases}
				 |q_{evi}(t)|+\sigma_{\rho}-\rho_{ni}(t),\\
				 \quad\text{for}\quad|q_{evi}(t)|+\sigma_{\rho}-\rho_{ni}(t)\ge \sigma_{s}\\
				 \sigma_{s},\quad\text{otherwise}
		\end{cases}
	\end{equation}
	where $\sigma_{\rho}>0$ and $\sigma_{s}>0$ are proper large enough constants. It can be observed that the designed projection operator has a time-varying upper bound, which is an extension of the conventional one with a constant upper bound.
	\begin{remark}
		It can be observed that $\rho^{i}_{\max}(t)$ is a piece-wisely smooth differentiable function. However, its time-derivative does not exists for $|q_{evi}(t)|+\sigma_{\rho}-\rho_{ni}(t)= \sigma_{s}$. Nevertheless, note that $\rho^{i}_{\max}(t)$ is Lipschitz continuous over the entire domain, therefore this rules out the possibility of $|\dot{\rho}^{i}_{\max}(t)|\to+\infty$.
	\end{remark}
	
	Subsequently,
	$\nabla^{i}_{\rho}(t)$ denotes the partial-derivative of $f(\Delta\rho_{i}(t),\rho^{i}_{\max})$ with respect to $\Delta\rho_{i}(t)$, expressed as $\nabla^{i}_{\rho}(t) = \frac{\partial f(\Delta\rho_{i}(t),\rho^{i}_{\max}(t))}{\partial\Delta\rho_{i}(t)}$, while $\nabla^{i}_{\rho\max}(t)$ is specified as $\frac{f(\Delta\rho_{i}(t),\rho^{i}_{\max}(t))}{\partial\rho^{i}_{\max}(t)}$, $\varpi_{i}(t)(i = 1,2,3)$ stands for the $i$-th component of $\boldsymbol{\varpi}$, $C_{\rho} > 0$ is a damping coefficient for the dynamical system, $C_{v} > 0$, $F_{v} > 0$ are two positive gain parameters that are used to control the excitation signal.

	 \begin{remark}\label{remarkPOD}
	 	For the given dynamical system, taking the time-derivative of the evaluation function $f(\Delta\rho_{i}(t),\rho^{i}_{\max}(t))$, it yields: $\dot{f}(\Delta\rho_{i}(t),\rho^{i}_{\max}(t)) 
	 			= \nabla^{i}_{\rho}(t)\left[1-f\right]-\frac{\partial f}{\partial \rho^{i}_{\max}}\dot{\rho}^{i}_{\max}(t)+\frac{\partial f}{\partial\rho^{i}_{\max}}\dot{\rho}^{i}_{\max}(t)$.
	 			
	 	Specifically, for $f(\Delta\rho_{i}(t),\rho^{i}_{\max}(t)) = 1$, we have $\dot{f}=0$, which guarantees that $\rho_{i}(t) < |q_{evi}(t)|+\sigma_{\rho}$ will be always hold. 
	 	Compared with other existing excitation dynamics \cite{yong2020flexible,mehdifar2022funnel}, the utilization of such a dynamic projection operator offers two notable advantages. Firstly, it allows for the choosing of the excitation gain parameter, denoted as $C_{v}$, to be sufficiently large value, thereby ensuring that the performance envelope can expand rapid enough once the contradiction is detected, thus the performance envelope constraint can be still remain satisfied. Additionally, this approach ensures that a large value of $C_{v}$ does not result in an excessively substantial modification signal, thereby guaranteeing $\Delta\rho_{i}(t)$ rapidly vanishes once the contradiction is resolved.
	 \end{remark}

	\subsection{Discussion on the Working Mechanism of the Compatible Performance Control Scheme}
	In this subsection, we elaborate on the mechanism of the Compatible Performance Control framework (CPC), which can be concluded as two steps.
	
	\textbf{Step 1.} If the nominal performance function $\rho_{ni}(t)$ is not compatible with the angular velocity limitation, the trajectory of $q_{evi}^{2}(t)$ will approach $\rho_{ni}^{2}(t)$ rapidly, even with a saturated $\omega_{vi}(t)$. Therefore, this result in a decreasing of $\mathcal{D}_{i}(t)$ and $\beta_{i}(t)$, which further switches the indicator function $\mathcal{S}(\beta_{i}(t))$ from $0$ to $1$.
	Meanwhile, due to the saturation of $\omega_{vi}(t)$, a deviation occurs and leads to the nonzero of $|\varpi_{i}(t)|$, i.e., $|\varpi_{i}(t)| > 0$. With the previously designed $F_{v}$ and $C_{v}$, $\tanh(F_{v}\varpi_{i}(t))$ shows enough saturated output.
	As highlighted in Remark \ref{remarkPOD}, this generates a modification signal $\Delta\rho_{i}(t)$ that rapidly expands the  lumped performance envelope $\rho_{i}(t) > \rho_{ni}(t)$, therefore alleviates the contradiction and ensures that the state trajectory is able to keep within the post-modified performance envelope while still keeping the angular velocity limitation.
	
	\textbf{Step 2.} Once the nominal performance function $\rho_{ni}(t)$ is compatible with the angular velocity limitation, the indicator function $\mathcal{S}(\beta_{i}(t))$ will switch back to $0$. This implies that the dynamical system given in equation (\ref{modified}) becomes a simple converging form as $\Delta\dot{\rho}_{i}(t) = -C_{\rho}\Delta\rho_{i}(t)$. Therefore, $\Delta\rho_{i}(t)$ exponentially converges, ensuring that the total performance function $\rho_{i}(t)$ returns to the nominal one $\rho_{ni}(t)$, and the system will be converged while satisfying the originally designed performance criteria.

\begin{remark}\label{nocontradiction}
		There may be scenarios where $\omega_{vi}(t)$ is saturated but $\mathcal{S}(\beta_{i}(t)) = 0$ holds. This circumstance occurs when $q_{evi}^{2}(t)$ is able to move away from $\rho^{2}_{ni}(t)$ with a saturated $\omega_{vi}(t)$, indicating that there is actually no contradiction exists. Therefore, the system's capability is enough to ensure that the performance envelope constraint will not be violated, showing that it unnecessary to require a modification.
	These three situations arise in different control scenarios, and it will be further elaborated and analyzed in Subsection \ref{FurtherCase}.
\end{remark}

	\section{ADAPTIVE COMPATIBLE PERFORMANCE CONTROLLER DESIGN}\label{adaptiveCPC}
In this section, based on the aforementioned CPC scheme, we propose an adaptive compatible performance controller (ACPC) to stabilize the system $\boldsymbol{z}_{1}\in\mathbb{R}^{3}$, $\boldsymbol{z}_{2}\in\mathbb{R}^{3}$ that given in equation (\ref{errorsub}) in the presence of parameter uncertainties while ensuring the satisfaction of all aforementioned constraints. 
In order to address the impact of parameter uncertainties and ensures the convergence of $\boldsymbol{z}_{2}$-subsystem, we employ a time-varying adaptive gain technique to establish a parameter adaptive law, facilitating the rapid convergence of the system while providing a better transient behavior.
% Additionally, to address input saturation, an auxiliary system is introduced to compensate for the perturbation resulting from saturation issues.

\begin{figure}[hbt!]
	\centering 
	\includegraphics[width=0.5\textwidth]{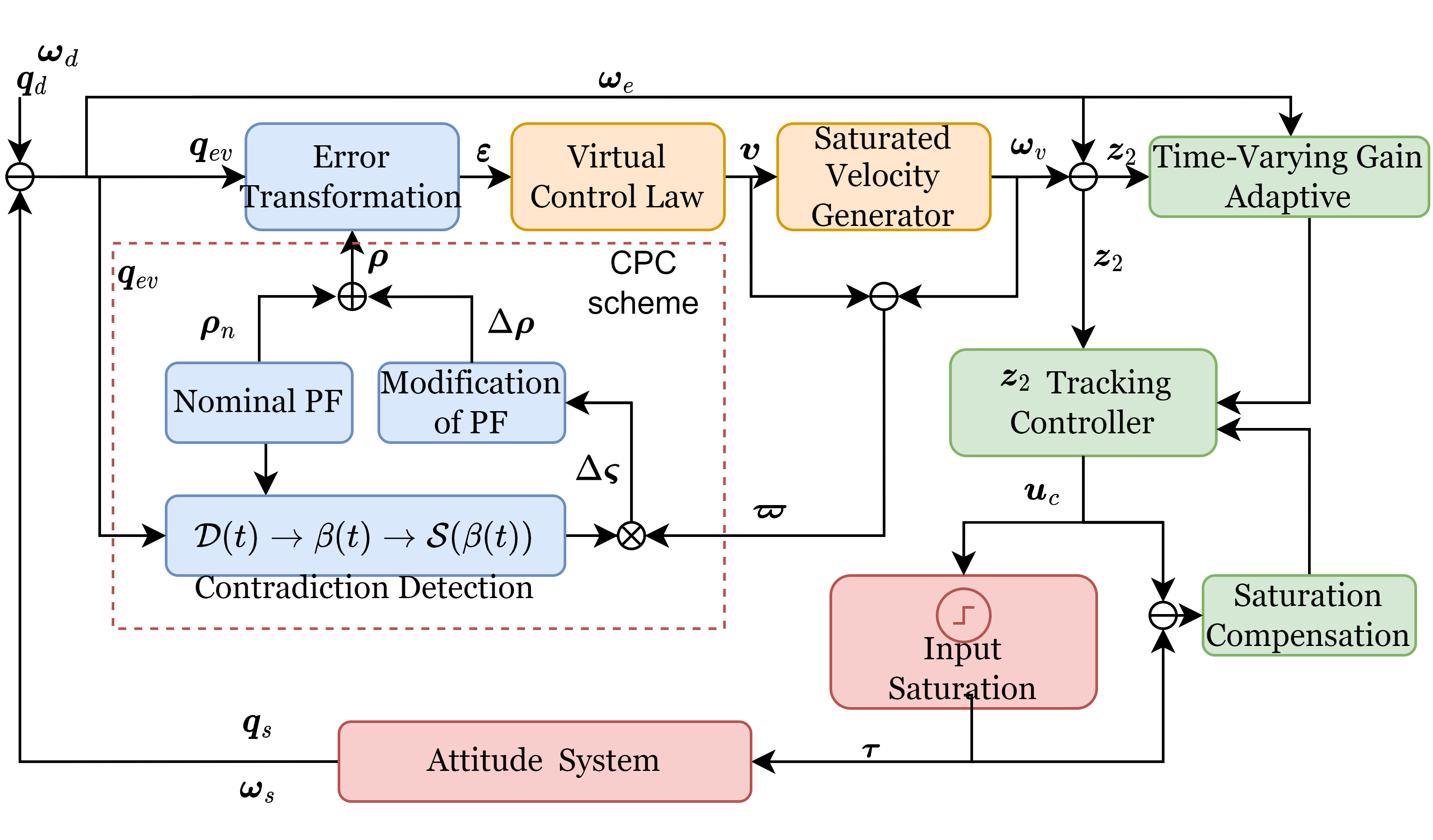}
	\caption{Main Structural Diagram of the Proposed Adaptive Compatible Performance Controller}       
	\label{controllerscheme}   
\end{figure}

A structural schematic diagram of the proposed adaptive CPC controller is depicted in Figure \ref{controllerscheme} for a further elaboration.
	
	Firstly, by applying the proposed bounded velocity generator (detailed in Subsection \ref{SVG}) and the performance function modification strategy (detailed in Subsection \ref{MODPPF}), it yields the following augmented system:
	\begin{equation}\label{closeloop2}
		\begin{aligned}
			\dot{\boldsymbol{\varepsilon}} &=
			\boldsymbol{\Psi}\boldsymbol{\varGamma}_{e}\boldsymbol{\omega}_{v} + \boldsymbol{\Psi}\boldsymbol{\varGamma}_{e}\boldsymbol{z}_{2} - \boldsymbol{\Lambda}\boldsymbol{\varepsilon}\\
			\Delta\dot{\boldsymbol{\rho}} &= \boldsymbol{G}_{\rho}\left[-C_{\rho}\Delta\boldsymbol{\rho} + C_{v}\Delta\boldsymbol{\varsigma}\right]-\boldsymbol{w} \\
			\dot{\boldsymbol{\omega}}_{v} &= -C_{d}C_{f}\boldsymbol{\omega}_{v} + C_{f}\boldsymbol{K}_{f}\boldsymbol{v}\\
			\boldsymbol{J}_{0}\dot{\boldsymbol{z}}_{2} &= \boldsymbol{W}\boldsymbol{\Theta} + \boldsymbol{\Omega}_{e} + \boldsymbol{u}_{c} + \Delta\boldsymbol{u} - \boldsymbol{J}_{0}\dot{\boldsymbol{\omega}}_{v}
		\end{aligned}
	\end{equation}
	where $\boldsymbol{G}_{\rho}\in\mathbb{R}^{3\times 3}$ is a diagonal matrix that defined as $\boldsymbol{G}_{\rho}\triangleq \text{diag}\left(g_{i}(t)\right)(i=1,2,3)$ (cf. equation (\ref{dotrhocompact})), $\boldsymbol{w}\in\mathbb{R}^{3\times1}$ is the element-spanned column vector of $\mathcal{W}_{i}(t)$, expressed as $\boldsymbol{w}\triangleq\text{vec}(\mathcal{W}_{i}(t))(i=1,2,3)$ (cf. equation (\ref{dotrhocompact})). $\Delta\boldsymbol{\varsigma}\in\mathbb{R}^{3}$ denotes the input of the performance modification dynamics, expressed as $\Delta\boldsymbol{\varsigma} \triangleq \text{vec}\left[\mathcal{S}(\beta_{i}(t))|\tanh(F_{v}\varpi_{i}(t))|\right](i=1,2,3)$; $\boldsymbol{K}_{f}\in\mathbb{R}^{3\times 3}$ denotes an element-spanned diagonal matrix, defined as $\boldsymbol{K}_{f} \triangleq \text{diag}(\mathcal{K}(\omega_{vi}(t),B_{\omega}))(i = 1,2,3)$.

%	Subsequently, in order to ensure that the translated variable $\boldsymbol{\varepsilon}$ is able to satisfy the performance envelope constraint, i.e., $|\varepsilon_{i}(t)| < 1$ holds for $\forall t\in \left[0,+\infty\right]$ with $|\varepsilon_{i}(0)| < 1$, we further introduce a Barrier Lyapunov function into the controller design, expressed as follows:
%	\begin{equation}
%		V_{B} = \frac{k_{B}}{2}\sum_{i = 1}^{3}\ln\left(\frac{1}{1-\varepsilon^{2}_{i}(t)}\right)
%	\end{equation}
%    where $k_{B}>0$ is a design parameter. Accordingly, taking the time-derivative of $V_{B}$, one can be obtained that:
%    \begin{equation}\label{Vb}
%    	\begin{aligned}
%    		\dot{V}_{B} &= \frac{k_{ B}}{2}\sum_{i=1}^{3}\left(1-\varepsilon^{2}_{i}(t)\right)\frac{2\varepsilon_{i}(t)\dot{\varepsilon}_{i}(t)}{\left[1-\varepsilon^{2}_{i}(t)\right]^{2}}\triangleq \boldsymbol{\varepsilon}^{\text{T}}\boldsymbol{G}_{\varepsilon}\dot{\boldsymbol{\varepsilon}}
%    	\end{aligned}
%    \end{equation}

\subsection{Main Design of the Adaptive Compatible Performance Controller}
    
    \textbf{1.} The input of the bounded velocity generator $\boldsymbol{v}\in\mathbb{R}^{3}$ is designed as follows:
    \begin{equation}\label{vdesign}
    	\begin{aligned}
    		\boldsymbol{v}  &= \frac{-1}{C_{f}}\boldsymbol{K}^{-1}_{f}\left[\boldsymbol{\varGamma}^{\text{T}}_{e}\left(\boldsymbol{\Psi}\boldsymbol{G}_{\varepsilon}\boldsymbol{\varepsilon}+K_{q}\boldsymbol{q}_{ev}\right)-\frac{\boldsymbol{\omega}_{v}\boldsymbol{\varepsilon}^{\text{T}}\boldsymbol{G}_{\varepsilon}\boldsymbol{\Lambda}\boldsymbol{\varepsilon}}{\|\boldsymbol{\omega}_{v}\|_{2}^{2}} - K_{\xi}\boldsymbol{\xi}\right]
    	\end{aligned}
    \end{equation}
    where $C_{f} > 0$ is a parameter of the bounded velocity generator, $K_{q}, K_{\xi} > 0$ are controller gain parameters,
    $\boldsymbol{G}_{\varepsilon}$ denotes a $\mathbb{R}^{3\times 3}$ diagonal matrix that defined as $\boldsymbol{G}_{\varepsilon} \triangleq \text{diag}\left(k_{B}\frac{1}{1-\varepsilon^{2}_{i}(t)}\right)(i = 1,2,3)$, with $k_{B} > 0$ is a positive design parameter. $\boldsymbol{\xi}\in\mathbb{R}^{3}$ is an auxiliary compensation system, specified as follows:
\begin{equation}\label{xi}
	\dot{\boldsymbol{\xi}} = -\left[p_{1}+p_{\xi}\frac{\|\Delta\boldsymbol{\varsigma}\|_{2}^{2}+\|\boldsymbol{w}\|_{2}^{2}+p_{c}/p_{\xi}}{\|\boldsymbol{\xi}\|_{2}^{2}}\right]\boldsymbol{\xi} + p_{2}\left(\Delta\boldsymbol{\varsigma}+\boldsymbol{w}\right)
\end{equation}
    where $p_{1}, p_{2}, p_{\xi} > 0$ are positive design parameters, $p_{c}$ is an additional constant that designed for robust consideration.
    
    \textbf{2.} The command actual control law $\boldsymbol{u}_{c}$ is designed as follows:
\begin{equation}\label{uc}
	\begin{aligned}
		\boldsymbol{u}_{c} = -\boldsymbol{W}\hat{\boldsymbol{\Theta}} + \boldsymbol{J}_{0}\dot{\boldsymbol{\omega}}_{v}-\boldsymbol{\Omega}_{e} + K_{\delta}\boldsymbol{\delta} - K_{2}\boldsymbol{J}_{0}\boldsymbol{z}_{2} + K_{\gamma}\boldsymbol{\gamma}_{\varepsilon}
	\end{aligned}
\end{equation}
where $K_{2}, K_{\delta}>0$ and $K_{\gamma}>0$ are constant gain parameters,
 $\boldsymbol{\delta}\in\mathbb{R}^{3}$ denotes the output of the anti-wind up dynamical compensation system, given as follows:
\begin{equation}\label{delta}
	\dot{\boldsymbol{\delta}} = -\left[n_{1}+n_{\delta}\frac{\|\Delta\boldsymbol{u}\|_{2}^{2}}{\|\boldsymbol{\delta}\|_{2}^{2}}\right]\boldsymbol{\delta} + n_{2}\Delta\boldsymbol{u}
\end{equation}
where $n_{1},n_{2} > 0$ and $n_{\delta} > 0$ are positive gain parameters. 
Similarly, $\boldsymbol{\gamma}_{\varepsilon}\in\mathbb{R}^{3}$ is generated by another anti-windup dynamical system, specified as follows:
\begin{equation}\label{gamma}
		\dot{\boldsymbol{\gamma}}_{\varepsilon} = -\left[g_{1}+g_{\gamma}\frac{\|\boldsymbol{z}_{2}\|^{2}_{2}\|\boldsymbol{P}_{\varepsilon}\|_{2}^{2}+g_{c}/g_{\gamma}}{\|\boldsymbol{\gamma
			}_{\varepsilon}\|_{2}^{2}}\right]\boldsymbol{\gamma
		}_{\varepsilon} + g_{2}\|\boldsymbol{P}_{\varepsilon}\|_{2}\boldsymbol{z}_{2}
\end{equation}
where $g_{1}, g_{2} > 0$ and $g_{\gamma} > 0$ are positive design parameters, $g_{c}>0$ is an additional robust design parameter. $\boldsymbol{P}_{\varepsilon}\in\mathbb{R}^{3}$ denotes a coupling-term that coming from the $\boldsymbol{z}_{1}$-subsystem, defined as $\boldsymbol{P}_{\varepsilon}\triangleq \boldsymbol{\varGamma}^{\text{T}}_{e}\left[\boldsymbol{G}_{\varepsilon}\boldsymbol{\Psi}\boldsymbol{\varepsilon}+K_{q}\boldsymbol{q}_{ev}\right]$.

 $\hat{\boldsymbol{\Theta}}\in\mathbb{R}^{9\times 1}$ denotes the estimation for the unknown parameter $\boldsymbol{\Theta}$.
Specifically, the estimation $\hat{\boldsymbol{\Theta}}$ is updated by a projection-operator governed adaptive law as:
\begin{equation}
	\dot{\hat{\boldsymbol{\Theta}}} \triangleq \text{Proj}(\hat{\boldsymbol{\Theta}},\zeta(t)\boldsymbol{W}^{\text{T}}\boldsymbol{z}_{2},f_{\Theta}(\hat{\boldsymbol{\Theta}}))
\end{equation}
where $\text{Proj}(\cdot,\cdot,\cdot)$ is previously defined in Definition \ref{Defproj}. Specifically, it can be expressed as:
	\begin{equation}\label{theta}
	\dot{\hat{\boldsymbol{\Theta}}} \triangleq 
	\begin{cases}
		\zeta(t)\boldsymbol{W}^{\text{T}}\boldsymbol{z}_{2}- \frac{\boldsymbol{\nabla}_{\Theta}\boldsymbol{\nabla}^{\text{T}}_{\Theta} \cdot f_{\Theta}(\hat{\boldsymbol{\Theta}})}{\|\boldsymbol{\nabla}_{\Theta}\|_{2}^{2}}\zeta(t)\boldsymbol{W}^{\text{T}}\boldsymbol{z}_{2},\\
		\text{for}\quad f_{\Theta}(\hat{\boldsymbol{\Theta}})>0 \text{and}\boldsymbol{\nabla}_{\Theta}^{\text{T}}\zeta(t)\boldsymbol{W}^{\text{T}}\boldsymbol{z}_{2}> 0\\
		\zeta(t)\boldsymbol{W}^{\text{T}}\boldsymbol{z}_{2},\quad\text{otherwise}
	\end{cases}
\end{equation}
where $f_{\Theta}(\hat{\boldsymbol{\Theta}})$ denotes the evaluation function corresponds to $\hat{\boldsymbol{\Theta}}$, which has the same form as defined in Definition \ref{Lfunc}, $\boldsymbol{\nabla}_{\Theta}\in\mathbb{R}^{9}$ denotes the element-spanned column vector that consisted of partial-derivatives of $f_{\Theta}(\hat{\boldsymbol{\Theta}})$ with respect to each $i$-th component of $\hat{\boldsymbol{\Theta}}(i=1,2...9)$, expressed as $\boldsymbol{\nabla}_{\Theta}=\left[\frac{\partial f(\boldsymbol{\hat{\Theta}})}{\partial \hat{\Theta}_{1}},...,\frac{\partial f(\boldsymbol{\hat{\Theta}})}{\partial \hat{\Theta}_{N}}\right]^{\text{T}}\in\mathbb{R}^{9}$.

$\zeta(t)\in\mathbb{R}$ is a scalar variable that denotes the time-varying adaptive gain, which is updated by another projection operator-governed dynamics, expressed as:
\begin{equation}
	\dot{\zeta}(t) \triangleq\lambda_{\zeta} \text{Proj}(\zeta(t),\mathcal{Z}(t),f_{\zeta}(\zeta(t)))
\end{equation}
where $\lambda_{\zeta} > 0$ is a constant gain parameter that needs indicating. Accordingly, $\dot{\zeta}(t)$ can be specified as follows:
\begin{equation}
	\begin{aligned}\label{zeta}
			\dot{\zeta}(t) &\triangleq
			\begin{cases}
				\lambda_{\zeta}\mathcal{Z}(t)\left[1-\frac{\nabla_{\zeta}\cdot\nabla_{\zeta}f_{\zeta}(\zeta(t))}{\nabla^{2}_{\zeta}}\right]\\
				\quad \text{for} f_{\zeta}(\zeta(t)) > 0\text{and}\nabla_{\zeta}\mathcal{Z}(t) > 0\\
				\lambda_{\zeta}\mathcal{Z}(t),\quad\text{otherwise}\\
			\end{cases}\\
			\mathcal{Z}(t) &\triangleq 2\left[\zeta(t) - K_{\zeta}\chi(t)\zeta^{2}(t)\right]\\
	\end{aligned}
\end{equation}
where $K_{\zeta} > 0$ is a gain parameter, $\nabla_{\zeta}\in\mathbb{R}$ denotes the partial-derivative of $f_{\zeta}(\zeta(t))$ with respect to $\zeta(t)$, i.e. $\nabla_{\zeta} = \frac{\partial f_{\zeta}(\zeta)}{\partial \zeta}$. To facilitate the following analysis, we further rearrange equation (\ref{zeta}) into a compact form by defining a factor $\mu(t)$ as
	$\mu
	(t) \triangleq \left[1-\frac{\nabla_{\zeta}\cdot\nabla_{\zeta}f_{\zeta}(\zeta(t))}{\nabla^{2}_{\zeta}}\right]\text{or}\mu(t) = 1$.
Thereby it concludes the compact form of the $\zeta$-dynamics:
\begin{equation}\label{zetacompact}
	\dot{\zeta}(t) = \lambda_{\zeta}\mu(t)\mathcal{Z}(t)
\end{equation}
\begin{remark}\label{mucharas}
	It can be observed that $\mu(t) = 0$ holds for $f_{\zeta}(\zeta(t)) = 1$, which indicate that $\zeta(t)$ is at the upper bound of its allowed varying region. Meanwhile, $\mu(t) = 1$ holds for all circumstances that correspond to the second condition in equation (\ref{zeta}), and $\mu(t)\in\left(0,1\right)$ holds for $f_{\zeta}(\zeta(t))\in\left(0,1\right)$.
\end{remark}
The variable $\chi(t)$ in equation (\ref{zeta}) is a time-varying damping coefficient that is generated by the following dynamics:
\begin{equation}\label{chi}
	\dot{\chi}(t) = -\lambda_{\chi}\chi(t) + \lambda_{\chi}\frac{\|\boldsymbol{W}\|_{2}^{2}}{1+\|\boldsymbol{W}\|^{2}_{2}}
\end{equation}
where $\lambda_{\chi}> 0$ is a positive constant damping coefficient.
\begin{remark}
		It should be noted that in our controller design, instead of adding a term expressed as $\boldsymbol{P}_{\varepsilon}$ on $\boldsymbol{u}_{c}$, we use a "filtered" $\boldsymbol{P}_{\varepsilon}$, i.e., $K\gamma\boldsymbol{\gamma}_{\varepsilon}$, to make such a compensation for $\boldsymbol{z}_{1}$-system. The main idea behind such a design is: for the simple close-loop system $\boldsymbol{z}_{2}$, the adding term $\boldsymbol{P}_{\varepsilon}$ actually behaves like a perturbation for $\boldsymbol{z}_{2}$-subsystem, which may lead to a worse transient behaviour. This problem will be especially prominent for controller designed with PPC scheme as the compensation term $\boldsymbol{P}_{\varepsilon}$ includes $\boldsymbol{G}_{\varepsilon}$ and $\boldsymbol{\Psi}$ may be potentially an extremely large value under some circumstances. However, it should be noticed that the essential purpose of adding the term $\boldsymbol{P}_{\varepsilon}$ is to cancel the backstepping-coupling term $\boldsymbol{z}^{\text{T}}_{2}\boldsymbol{P}_{\varepsilon}$, and such a term vanishes for $\|\boldsymbol{z}_{2}\|\to 0$ no matter how big $\|\boldsymbol{P}_{\varepsilon}\|$ is. Therefore, by leveraging an auxiliary system $\boldsymbol{\gamma}_{\varepsilon}$, the filter output considers on the value of $\|\boldsymbol{z}_{2}\|$ simultaneously.
\end{remark}

\subsection{Analysis on $\zeta(t)$ and $\chi(t)$}
This subsection discusses on some vital properties of the parameter adaptive system.

\begin{corollary}\label{coroChi}
$\chi(t)\in\left[0,1\right)$ holds for $\forall t\in\left[t_{0},+\infty
		\right)$ with the given initial condition that satisfies $\chi(t_{0}) < 1$.		
\end{corollary}

\begin{corollary}\label{coroZeta}
$\mu(t)\in\left[0,1\right]$, $\frac{1}{\zeta(t)}>0$ and $\frac{1}{\zeta(t)}\le\frac{1}{\zeta(t_{0})}+K_{\zeta}$ holds for $\forall t\in\left[t_{0},+\infty\right)$.
\end{corollary}

Proof of Corollary \ref{coroChi} and \ref{coroZeta} are detailed in Appendix \ref{ProofChi} and \ref{ProofZeta}.

	\section{STABILITY ANALYSIS}
	
	\subsection{Main Theorem}
	\begin{theorem}\label{T1}
		For the attitude system given in equation (\ref{attsys}), under the Assumption (\ref{KnownJ0})(\ref{unknownJ})(\ref{BoundedUnknown})(\ref{DisAss}), with the input of the bounded velocity generator $\boldsymbol{v}$ in equation (\ref{vdesign}), the actual control law $\boldsymbol{u}_{c}$ in equation (\ref{uc}), and the parameter adaptive system $\dot{\hat{\boldsymbol{\Theta}}}$, $\dot{\zeta}(t)$, $\dot{\chi}(t)$ given in equation (\ref{theta})(\ref{zeta})(\ref{chi}), the closed loop system will be asymptotically converged to a small residual set near the equilibrium point $\boldsymbol{q}_{ev} = \boldsymbol{0}$, $\boldsymbol{\omega}_{e} = \boldsymbol{0}$ under time-varying parameter uncertainties, and the closed-loop will be ultimately bounded. Meanwhile, all constraints that given in equation (\ref{angcons})(\ref{inputcons})(\ref{percons}) are satisfied during the whole control process. 
	\end{theorem}

The Proof of Theorem \ref{T1} is detailed in Appendix \ref{prooftheorem}.

\subsection{Parameter Selecting Suggestion}
	Except of trivial parameter selection suggestion on providing sufficiently large parameters to ensure that $C_{d}C_{f}-\frac{K_{\xi}}{2}$, $G_{2}$, $G_{P}-\frac{1}{2b_{1}}$, $G_{\xi}$, $G_{\varsigma}$, $G_{\delta}$, $G_{\gamma}$, $G_{w}$ and $G_{u}$ are all positive (See Appendix \ref{prooftheorem} for detailed elaboration), this subsection will provide additional detailed suggestions for constructing the CPC scheme.
	
Firstly, for the proposed bounded velocity generator in Subsection \ref{SVG}, it can be noted that a larger parameter $C_{d}$ will lead to a more "conservative" output, that is, the output of $\omega_{vi}(t)$ may not make full use of its allowed maximum value $B_{\omega}$. However, according to Remark \ref{Bw}, an appropriately large $C_{d}$ should better be chosen, which will yield a smaller $\|\dot{\boldsymbol{\omega}}_{d}\|_{2}$ and thus provide a smoother attitude transition process. 

Secondly, for the indicator function $\mathcal{S}(\beta_{i}(t))$ designed in Subsection \ref{detection}, $\epsilon$ can be properly chosen to be positive to render an in-advance contradiction detection, such as $\epsilon = 0.1$. Meanwhile, for the dynamic projection operator-based modification dynamics $\Delta\dot{\rho}_{i}(t)$, the choosing of $C_{v}$ should be sufficiently large to obtain a rapid expanding when such a contradiction exists. As we highlighted in Remark \ref{remarkPOD}, this utilization of projection operator ensures that the post-modified performance envelope will not be too far away from the state trajectory, therefore this provide a more flexible choosing of $C_{v}$.
Although that the satisfaction of the angular velocity limitation is based on the convergence of $\boldsymbol{z}_{2}$, nevertheless, the impact of nonzero $\boldsymbol{z}_{2}$ can be alleviated by providing additional margin when selecting the parameter $B_{\omega}$.

	\section{NUMERICAL SIMULATION AND ANALYSIS}\label{CPCsimulation}
	In this section, we presents several groups of simulation results to validate the effectiveness of the proposed adaptive compatible performance controller (ACPC).
%	 Firstly, we provide several attitude control scenarios with randomly provided initial conditions to show the basic effect of the proposed scheme, and we then discuss on the effectiveness of the adaptive strategy, discussing its effect on addressing disturbance and parameter uncertainties.
%	Further, through a comparison simulation, we focus on the effectiveness of the contradiction detection and alleviation strategy, further showing that the stated incompatibility problem indeed lead to an infeasibility of the system as we analyzed in Subsection \ref{conflict}, and the proposed CPC scheme is a proper solution to this issue.

In this subsection, the spacecraft is assumed to be a rigid-body spacecraft with fast-varying unknown inertia parameters. The known nominal inertia matrix is given as $\boldsymbol{J}_{0} = \text{diag}\left(2,2,2\right)(kg\cdot m^{2})$, while the time-varying unknown part $\Delta\boldsymbol{J}\in\mathbb{R}^{3\times 3}$ is given as:
\begin{equation}
	\Delta\boldsymbol{J} = 
	\begin{bmatrix}
		\Delta_{J11}&\Delta_{J12}&\Delta_{J13}\\
			\Delta_{J12}&\Delta_{J22}&\Delta_{J23}\\
				\Delta_{J13}&\Delta_{J23}&\Delta_{J33}\\
	\end{bmatrix}\in\mathbb{R}^{3\times 3}
\end{equation}
with each element $\Delta_{J11}$, $\Delta_{J12}$, $\Delta_{J13}$, $\Delta_{J22}$, $\Delta_{J23}$, $\Delta_{J33}$ are provided as follows.
\begin{equation}
	\begin{cases}
		\Delta_{J11} = 
		0.2\cdot\sin(0.1t+20)+0.3\cdot \cos(0.2t+10)\\
		\Delta_{J12} = 
		0.1\cdot \sin(0.6t+10)+0.2\cdot \cos(0.2t+10)+0.2\\
		\Delta_{J13} = 
		0.1\cdot \sin(0.5t)+0.1\cdot \cos(0.2t+10)+0.2\\
		\Delta_{J22} = 
		0.3\cdot \sin(0.5t)+0.3\cdot \cos(0.2t+10)\\
		\Delta_{J23} = 
		0.1\cdot \sin(0.3t+20)+0.1\cdot \cos(0.2t+10)+0.2\\
		\Delta_{J33} = 
		0.4\cdot \sin(0.2t)+0.3\cdot \cos(0.2t+10)\\	
	\end{cases}
\end{equation}
Accordingly, the total inertia matrix is given as $\boldsymbol{J} = \boldsymbol{J}_{0} + \Delta\boldsymbol{J}$.
Subsequently, the external environmental disturbance $\boldsymbol{d}_{e}$ is exerted as follows:
\begin{equation}
	\boldsymbol{d}_{e} = 
	\begin{bmatrix}
		1e-3 \cdot \left[4\sin\left(3\omega_{p}t\right) + 3\cos\left(10\omega_{p}t\right) -40\right]\\ 	
		1e-3\left[-1.5\sin\left(2\omega_{p}t\right) + 3\cos\left(5\omega_{p}t\right) +40\right]\\ 	
		1e-3\left[3\sin\left(10\omega_{p}t\right) - 8\cos\left(4\omega_{p}t\right) +45\right]\\ 	
	\end{bmatrix}
\end{equation}
where $\omega_{p} = 0.01$. It can be observed that the maximum external disturbance is about to be $0.02N\cdot m$, which is much larger than a general one in actual space environment, and thus is enough for the validation of the proposed scheme's robustness.
The maximum output ability of the spacecraft's actuator is set to be $0.05N\cdot m$.

Additionally, each simulation is performed at a step of $0.1s$, and the simulation time is $200s$. Main parameters that used in this simulation campaign are listed as follows: $C_{d} = 1$, $C_{f} = 4$, $\alpha = 1$, $\epsilon=0.001$, $\sigma_{s} = 0.1$, $C_{\rho} = 0.2$, $C_{v} = 10$, $F_{v} = 8$, $K_{2} = 10$. Main parameters for adaptive strategy is $\lambda_{\zeta} = 0.2$, $K_{\zeta} = 0.2$, $\lambda_{\chi} = 0.01$, $\zeta(t_{0}) = 2$, $\zeta_{\max} = 30$. The main parameter for nominal performance function is: $\rho_{0i} = 1$, $\rho_{\infty i} = 5\cdot 10^{-3}$, $\gamma_{i} = 0.08$.

\subsection{Fundamental Case: an attitude tracking control scenario}\label{case1}
We first provide a simulation case of an attitude tracking control scenario. Specifically, main constraints are given as follows:
\textbf{1.} The maximum angular velocity should not exceed $1.5^{\circ}/s$, therefore $\Omega_{\max} = 0.0262\text{rad}/s$ holds, such that $|\omega_{si}(t)| < \Omega_{\max}$ should be always respected.
\textbf{2.} The system should satisfy a control accuracy that higher than $0.02^{\circ}$, which can be approximately considered as $\max\left(|q_{evi}(t)|\right) < 9e-5$. 
The initial condition $\boldsymbol{q}_{s}(t_{0})$ and $\boldsymbol{\omega}_{s}(t_{0})$ are randomly selected as: $\boldsymbol{q}_{s}(t_{0})=\left[-0.432,0.34,-0.696,-0.459\right]^{\text{T}}$ and $\boldsymbol{\omega}_{s}(t_{0}) = \boldsymbol{0}$.
The target attitude quaternion $\boldsymbol{q}_{d}(t)$ is determined by the initial condition of $\boldsymbol{q}_{d}(t)$, i.e., $\boldsymbol{q}_{d}(t_{0})$, and then propagated by $\boldsymbol{\omega}_{d}(t)$, specified as follows:
\begin{equation}
	\begin{aligned}
			\boldsymbol{q}_{d}(t_{0})&=[0,0,0,1]^{\text{T}},
			\boldsymbol{\omega}_{d}(t)=0.5\left[\cos\frac{t}{c_{1}},\sin\frac{t}{c_{2}},-\cos\frac{t}{c_{3}}\right]^{\text{T}}
	\end{aligned}
\end{equation}
where $c_{1} = 80$, $c_{2} = 150$, $c_{3} = 100$. The unit is $^{\circ}/s$.

\begin{figure}[hbt!]
	\centering 
	\includegraphics[width=0.5\textwidth]{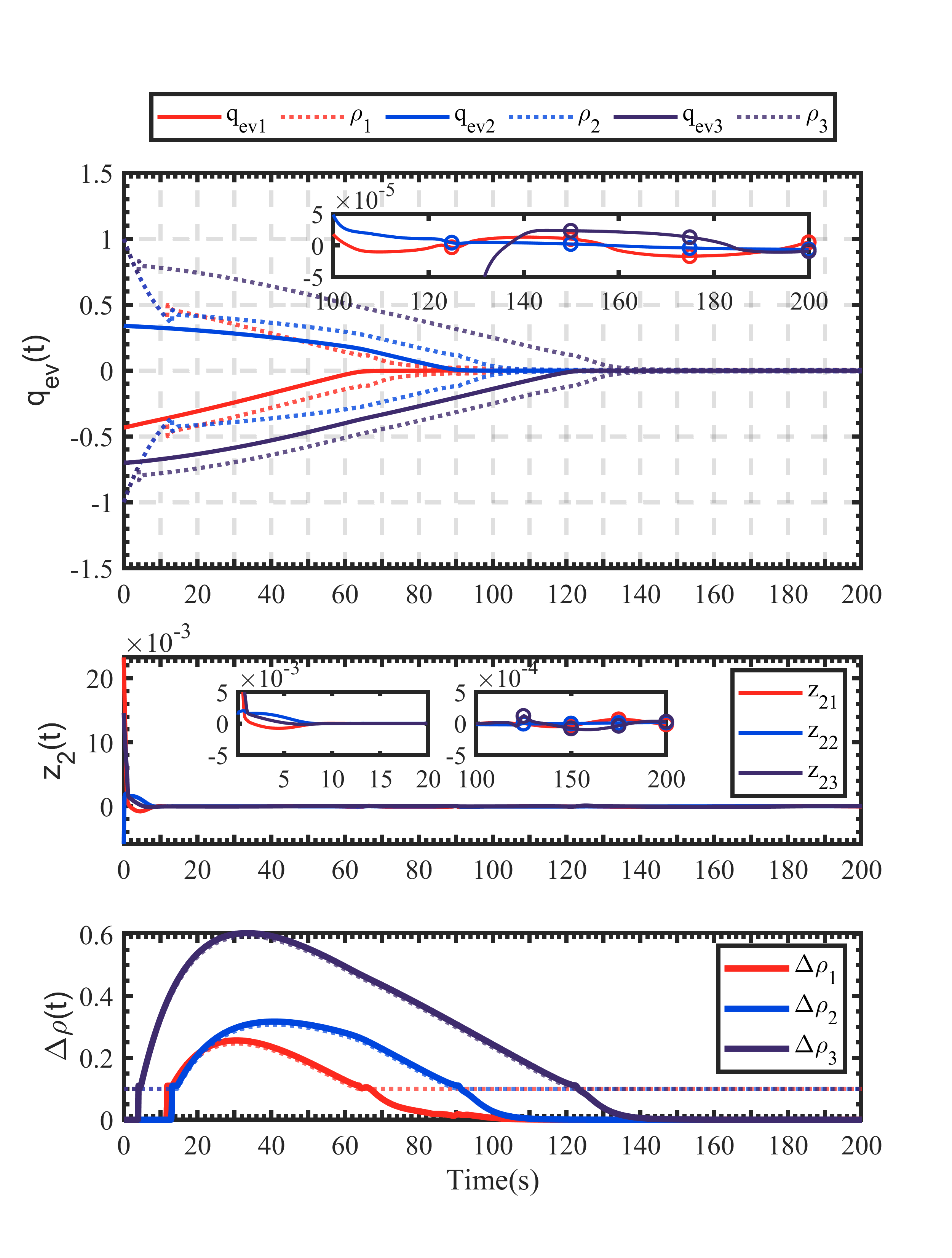}
	\caption{Time-Evolution of the error quaternion $\boldsymbol{q}_{ev}$, the Backstepping Tracking Error $\boldsymbol{z}_{2}$ and the Performance Function Modification Signal $\Delta\boldsymbol{\rho}$ (Fundamental Case)}       
	\label{QE1}  
\end{figure}
Figure \ref{QE1} illustrates the time-evolution of the vector part of the attitude error quaternion $\boldsymbol{q}_{ev}(t)$, the performance function modification signal $\Delta\rho_{i}(t)$ and the  Backstepping tracking error variable $\boldsymbol{z}_{2}(t)$. In the first subfigure of Figure \ref{QE1}, the solid line denotes the $q_{evi}(t)$-trajectory, while the dotted line stand for the performance function $\rho_{i}(t)$ that corresponds to each $i$-th component. In the thrid subfigure of Figure \ref{QE1}, the dotted line (almost overlapped) denotes that upper bound $\rho^{i}_{\max}$ for $\Delta\rho_{i}(t)$ such that $\Delta\rho_{i}(t) < \rho^{i}_{\max}(t)$ should be satisfied.

\begin{figure}[hbt!]
 				\centering 
 \includegraphics[width=0.5\textwidth]{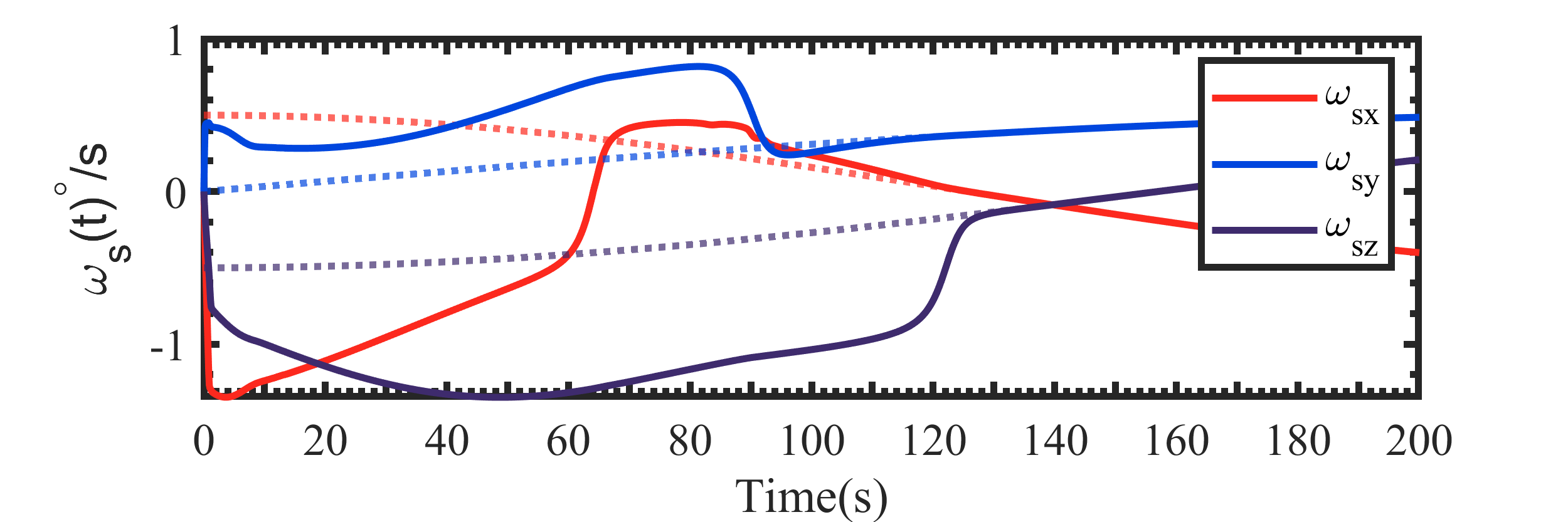}
 \caption{Time-Evolution of the angular velocity $\boldsymbol{\omega}_{s}$ (Fundamental Case)}       
 \label{WS1} 
 \centering 
 \includegraphics[width=0.5\textwidth]{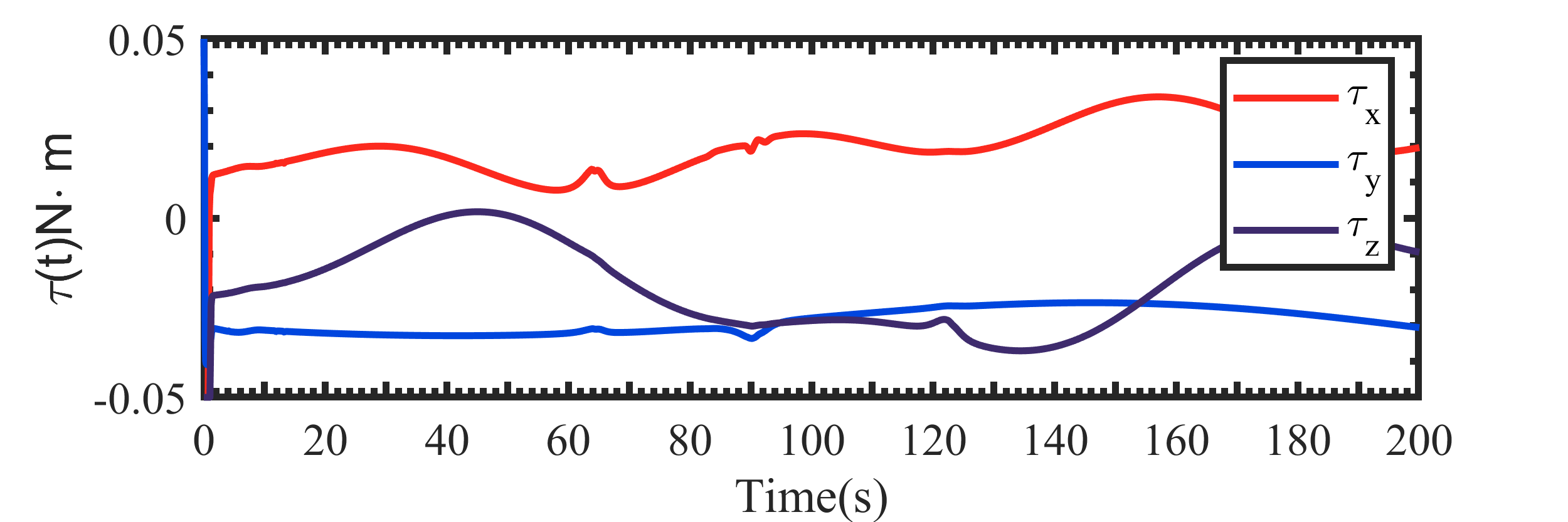}
 \caption{Time-Evolution of the Controller Output $\boldsymbol{\tau}$ (Fundamental Case)}       
 \label{TAU1}   
\end{figure}

Figure \ref{WS1} illustrates the time-evolution of the spacecraft's angular velocity $\boldsymbol{\omega}_{s}(t)$ that expressed in body-fixed frame $\mathfrak{R}_{b}$, while the dotted line stands for the target angular velocity $\boldsymbol{\omega}_{d}(t)$. Figure \ref{TAU1} shows the time-evolution of the actual exerted controller output $\boldsymbol{\tau}(t)$.

From Figure \ref{QE1}, it can be observed that at the first $5s$ of the simulation, $\Delta{\rho}_{i}(t) = 0$ and $\rho_{i}(t)$ is converging exponentially. However, note that owing to the given angular velocity constraint $\omega_{si}(t)<\Omega_{\max} = 0.0262$, the state trajectory $q_{evi}(t)$ can only converges slowly even if the spacecraft rotates at its maximum allowable angular velocity. Consequently, this triggers the designed contradiction detection strategy. It can be observed that $\rho_{i}(t)$ is immediately expanded by the modification signal $\Delta\rho_{i}(t)$, as illustrated in the first subfigure of Figure \ref{QE1}, and this alleviates the contradiction. Meanwhile, it can be discovered that $\rho_{i}(t)$ finally recovers to the nominal one $\rho_{ni}(t)$ such that $\Delta\rho_{i}(t)\to0$.
 Notably, since we have chosen $\sigma_{\rho} = 0.1$, thus the post-modified performance function satisfies $\rho_{i}(t)<|q_{evi}(t)|+0.1$, indicating that such a "modification" is restricted by the given varying upper bound, also validated by the simulation result in third subfigure of Figure \ref{QE1}.

For the satisfaction of constraints, it can be observed that $q_{evi}(t)$ finally converge to satisfy $|q_{evi}(t)| < 5e-5$, indicating that the system shows a high accuracy, and the performance requirements is satisfied. Simultaneously, as shown in Figure \ref{WS1}, it can be observed that $|\omega_{si}(t)|$ is strictly below the upper bound, indicating that the angular velocity limitation is satisfied.

\subsection{Further Validation on Handling Different Angular Velocity Limitation}\label{FurtherCase}
In this subsection, we present several groups of attitude reorientation control scenario to validate the effectiveness of the proposed scheme on handling different angular velocity limitation, i.e., $\boldsymbol{\omega}_{d} = \boldsymbol{0}$.

The initial condition is selected randomly as follows:
\begin{equation}
	\boldsymbol{q}_{s}(t_{0}) = \left[0.552,-0.627,0.200,-0.510\right]^{\text{T}},\boldsymbol{\omega}_{s}(t_{0}) = \boldsymbol{0}
\end{equation}

The target attitude quaternion is selected as $\boldsymbol{q}_{d} = \left[0,0,0,1\right]^{\text{T}}$, $\boldsymbol{\omega}_{d} = \boldsymbol{0}$. In this simulation, two different maximum angular velocity limitation is considered as: $\Omega_{\max} = 0.0175\text{rad}/s$ and $\Omega_{\max} = 0.0873\text{rad}/s$, equivalent to $1^{\circ}/s$ and $5^{\circ}/s$. The performance requirement is the same as in Subsection \ref{case1} to be $0.02^{\circ}$.

\begin{figure}[hbt!]
	\centering 
	\includegraphics[width=0.5\textwidth]{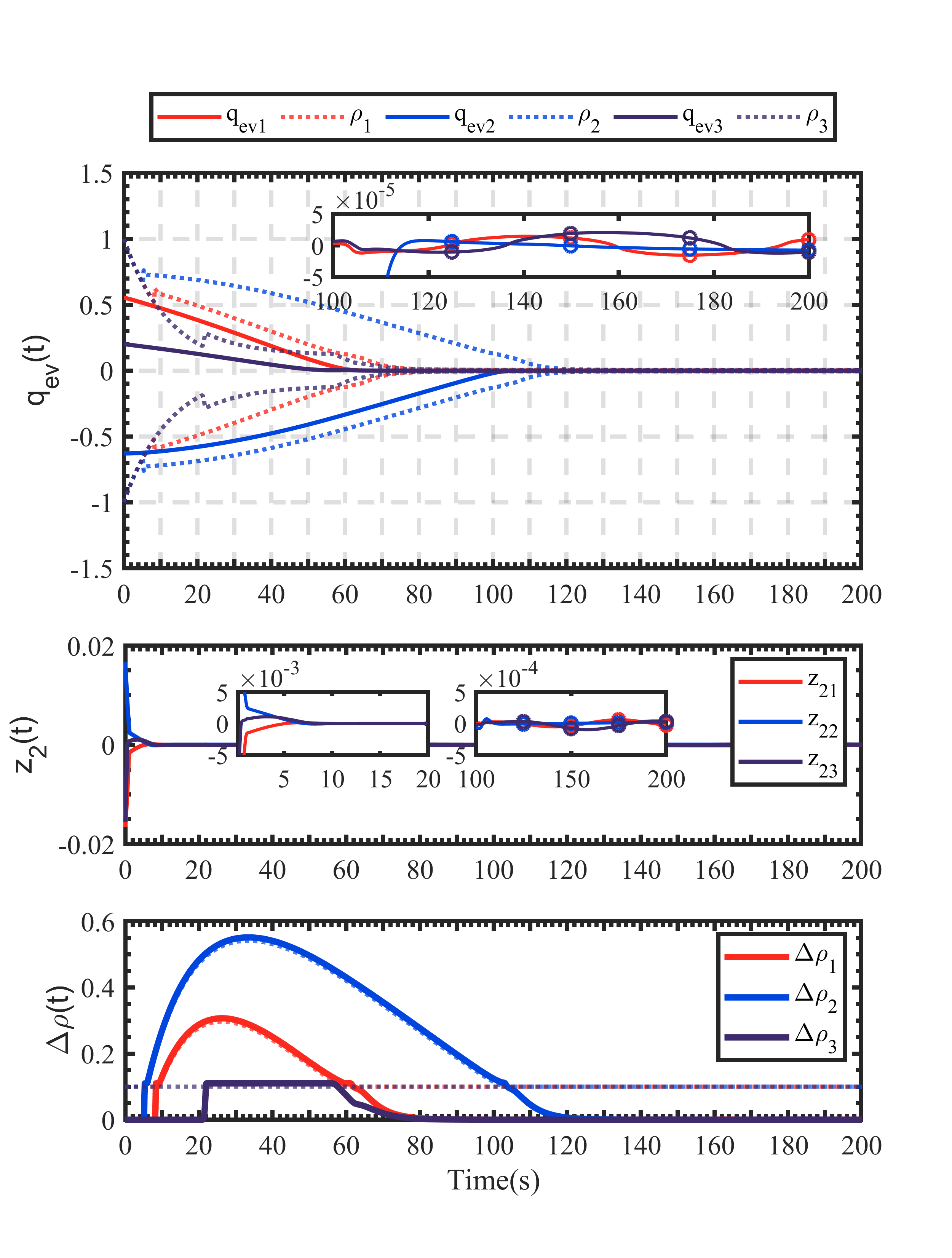}
	\caption{Time-Evolution of the error quaternion $\boldsymbol{q}_{ev}$, the Backstepping Tracking Error $\boldsymbol{z}_{2}$ and the Performance Function Modification Signal $\Delta\boldsymbol{\rho}$ (with $\Omega_{\max} = 0.0175\text{rad}/s$, equivalent to $1^{\circ}/s$)}       
	\label{QE2}    
		\centering 
	\includegraphics[width=0.5\textwidth]{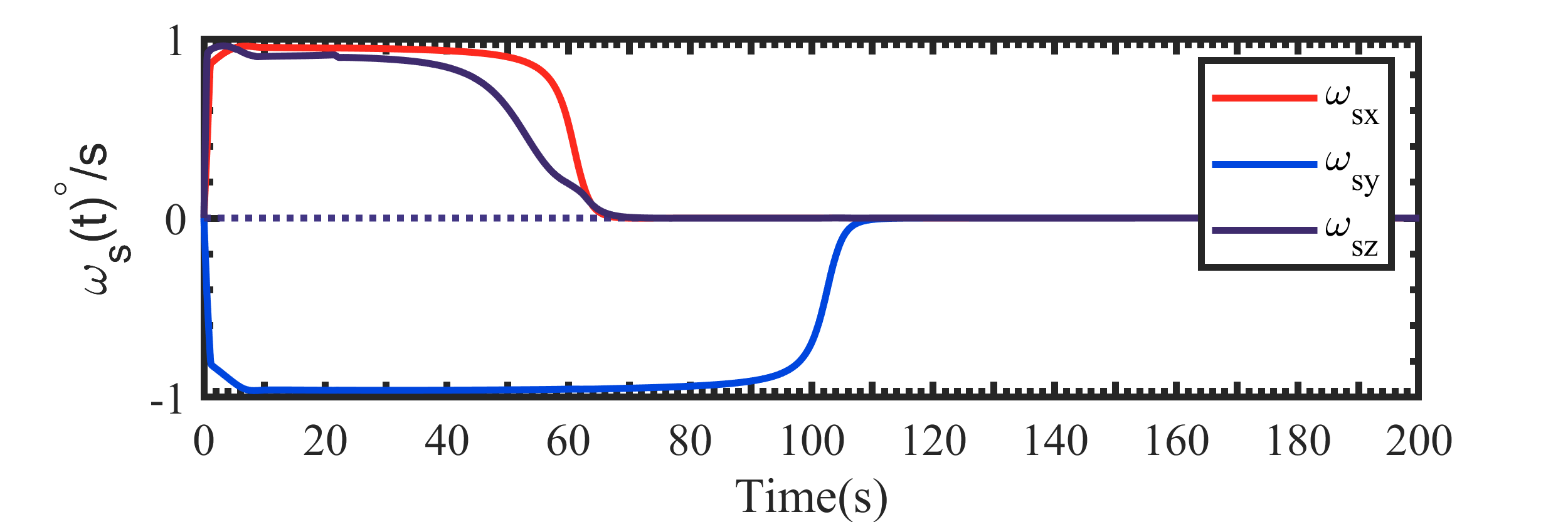}
	\caption{Time-Evolution of the angular velocity $\boldsymbol{\omega}_{s}$ (with $\Omega_{\max} = 0.0175$, equivalent to $1^{\circ}/s$)}       
	\label{WS2}  
\end{figure}
%\begin{figure}[hbt!]
%
%\end{figure}

\begin{figure}[hbt!]
	\centering 
	\includegraphics[width=0.5\textwidth]{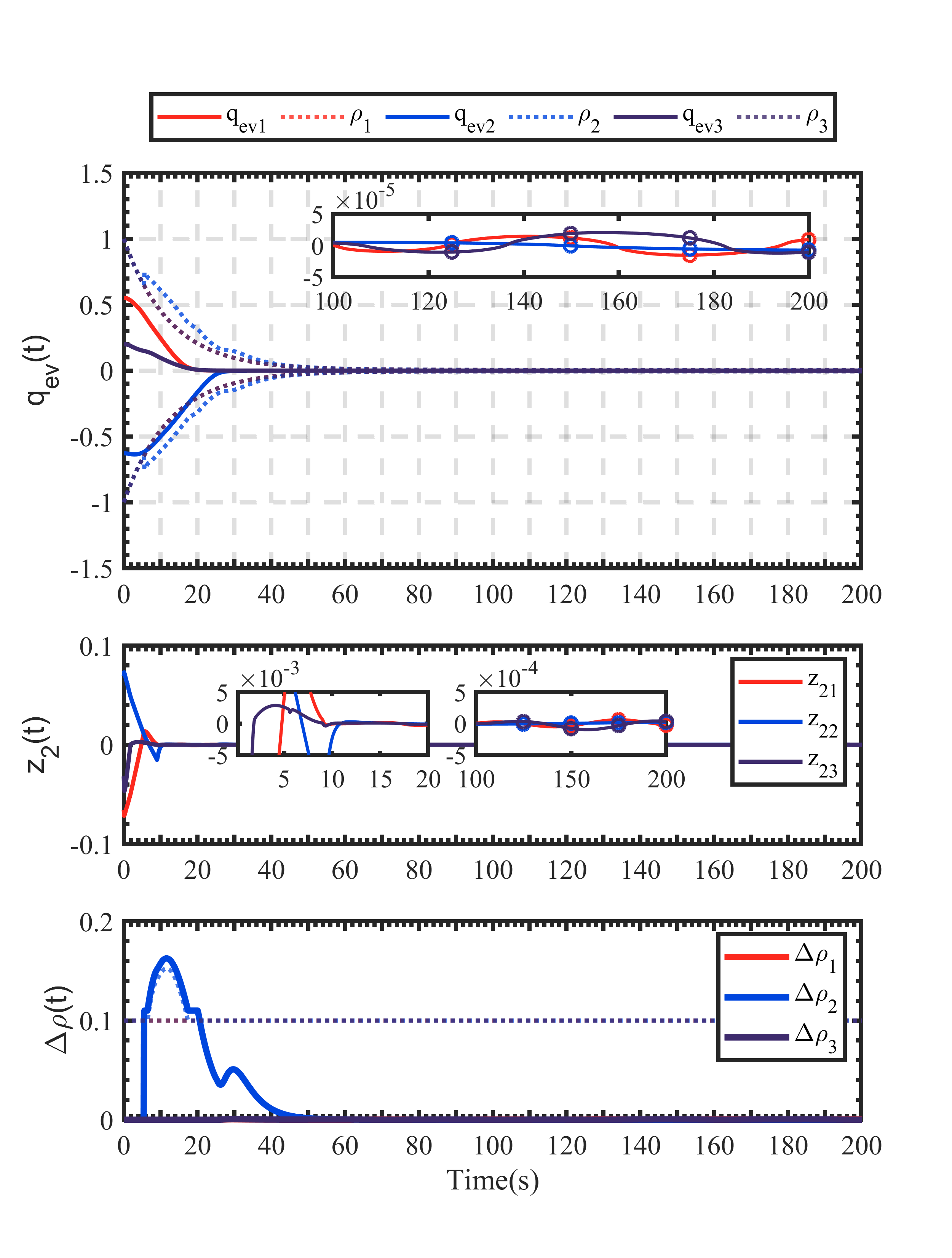}
	\caption{Time-Evolution of the error quaternion $\boldsymbol{q}_{ev}$, the Backstepping Tracking Error $\boldsymbol{z}_{2}$ and the Performance Function Modification Signal $\Delta\boldsymbol{\rho}$ (with $\Omega_{\max} = 0.0873$, equivalent to $5^{\circ}/s$)}       
	\label{QE3}     
\end{figure}

\begin{figure}[hbt!]
		\centering 
	\includegraphics[width=0.5\textwidth]{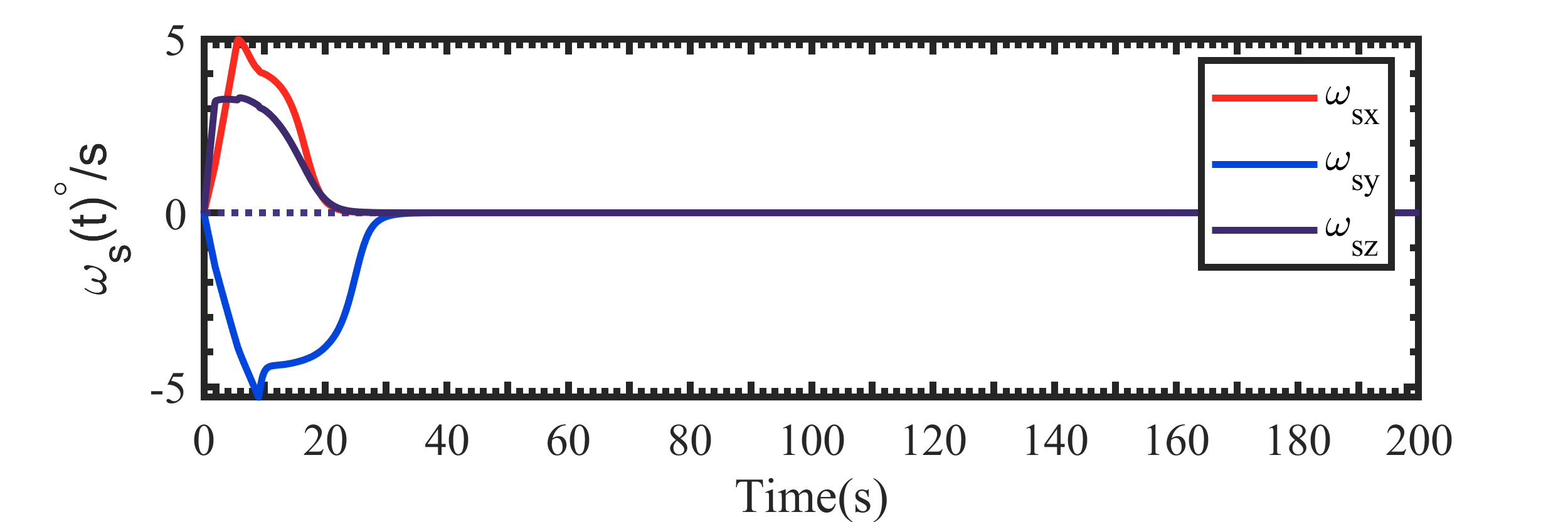}
	\caption{Time-Evolution of the angular velocity $\boldsymbol{\omega}_{s}$ (with $\Omega_{\max} = 0.0873$, equivalent to $5^{\circ}/s$)}       
	\label{WS3} 
\end{figure}

Figure \ref{QE2}, \ref{QE3} shows the time-evolution of $\boldsymbol{q}_{ev}(t)$, $\boldsymbol{z}_{2}(t)$ and $\Delta\boldsymbol{\rho}(t)$ with $\Omega_{\max} = 0.0175\text{rad}/s$ and $\Omega_{\max} = 0.0873\text{rad}/s$, respectively. It can be observed that $q_{evi}(t)$-trajectory is able to keep inside the performance envelope in all cases, and they both achieve an accuracy of $\max\left(|q_{evi}(t)|\right) < 2e-5$, indicating that the performance requirement is satisfied. Meanwhile, Figure \ref{WS2}, \ref{WS3} shows the time-evolution of $\boldsymbol{\omega}_{s}(t)$ during the whole control process, corresponding to $\Omega_{\max} = 0.0175\text{rad}/s$ and $\Omega_{\max} = 0.0873\text{rad}/s$, respectively. It can be observed that $\boldsymbol{\omega}_{s}(t)$ strictly satisfies the angular velocity limitation that given accordingly.
Notably, from Figure \ref{QE3}, it can be observed that there' s no modification sign generated for the first and the third component, i.e., $\Delta\rho_{1}(t) = 0$ and $\Delta\rho_{3}(t) = 0$. This can be explained as the given angular velocity allows the state trajectory to converge fast enough such that there's actually no contradiction exists. Therefore, this validates the effect of the indicator function $\mathcal{S}(\beta_{is})$ and our analysis in Remark \ref{nocontradiction}.

\subsection{Effectiveness of the Time-Varying Gain Adaptive}
In this subsection, we discuss on the effectiveness of the time-varying adaptive gain strategy.

We first introduce a constant version of the proposed controller, i.e., $\dot{\zeta}(t) = 0$, to make a comparison. It should be noticed this constant version adaptive have been utilized in many existing works, referring to \cite{hu2018adaptive}. Two different constant adaptive gain are selected as $\zeta_{1} = 5$ and $\zeta_{2} = 30$, denoted as \textit{CSmall} and \textit{CBig}.  
\begin{figure}[hbt!]
	\centering 
	\includegraphics[width=0.5\textwidth]{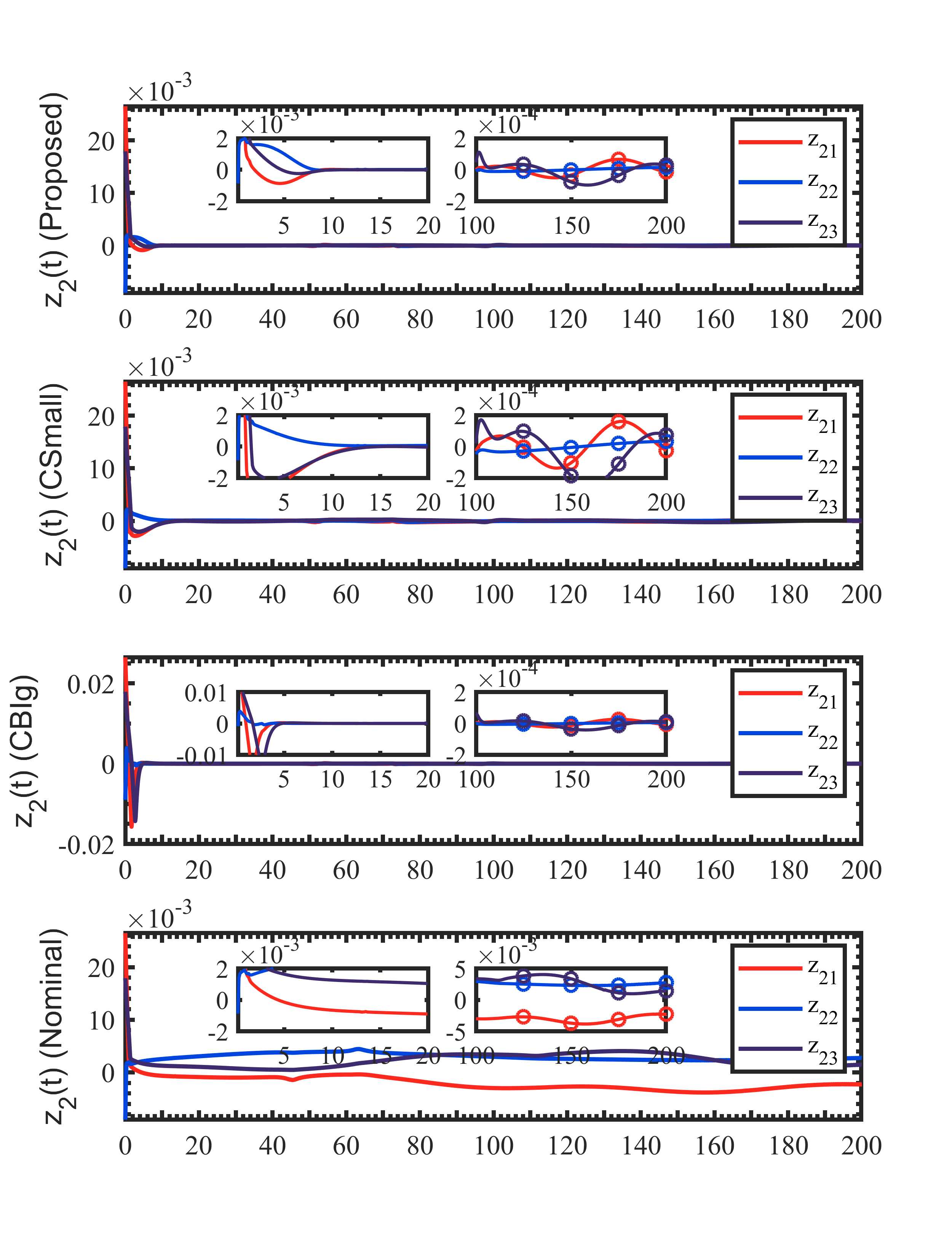}
	\caption{Time-Evolution of $\boldsymbol{z}_{2}(t)$ with proposed ACPC Controller, Benchmark \textit{CSmall}, \textit{CBig} and \textit{Nominal} Controller}       
	\label{ZCOMPARE}    
\end{figure}

Further, we introduce a non-uncertainty version of the proposed controller such that the accurate value of $\Delta\boldsymbol{J}$ is directly provided for the controller, therefore both $\boldsymbol{\Omega}_{e}$ and $\boldsymbol{\Omega}_{J}$ can be directly compensated. The external disturbance is handled by using a common technique that adding a widely-utilized $\tanh$-like term related to $D_{m}$. This benchmark controller is denoted as \textit{Nominal}.

The simulation setting including initial condition, target attitude and constraints are the same as given in Subsection \ref{case1}. The angular velocity limitation and the performance requirements are given as $\Omega_{\max} = 0.262\text{rad}/s$ and $\le 0.02^{\circ}$, respectively.

Figure \ref{ZCOMPARE} shows the time-evolution of $\boldsymbol{z}_{2}(t)$. For benchmark controller \textit{CSmall} and \textit{CBig}, a big overshooting occurs as $\boldsymbol{z}_{2}(t)$ converges and approaches the equilibrium point. Notably, the overshooting is extremely prominent for \textit{CBig} controller since $\zeta = 30$ is a large parameter. The proposed controller however, with only an overshoot smaller than $1e-3$, indicating that it is almost non-overshooting.
On the other hand, compared with the \textit{Nominal} benchmark controller, the proposed controller also shows a higher control accuracy. This is can be explained as the adaptive strategy considers the impact of parameter uncertainty and perturbation simultaneously, thus it provides a more accurate compensate for disturbance rejection, which lead to a higher accuracy. From Figure \ref{ZCOMPARE}, note that $\boldsymbol{z}_{2}(t)$-trajectory is able to converge at first, however, owing to the existing disturbance $\boldsymbol{d}_{e}$, it finally fail to converge to a small enough region.

\subsection{Additional Simulations to Support Theoretical Analysis: Existing Contradiction Between Constraints}
In this subsection, we provide a simulation validation of the contradicted constraint problem. We employ the control scheme in \cite{golestani2022prescribed}, which is a control scheme that directly combines the traditional PPC scheme with angular velocity limitation, to be the benchmark controller, denoted as \textit{TradPPC}. The simulation scenario and constraints are the same as in Subsection \ref{case1}, such that $\omega_{si}(t) < 0.0262\text{rad}/s$ should be satisfied.

 Specifically, the performance function of the \textit{TradPPC} benchmark controller is set to be an exponentially-converged one, of which the parameter is given as $\rho_{0i} = 1$, $\gamma_{i} = 0.05$ and $\rho_{\infty i} = 5\cdot 10^{-3}$.
\begin{figure}[hbt!]
	\centering 
	\includegraphics[width=0.5\textwidth]{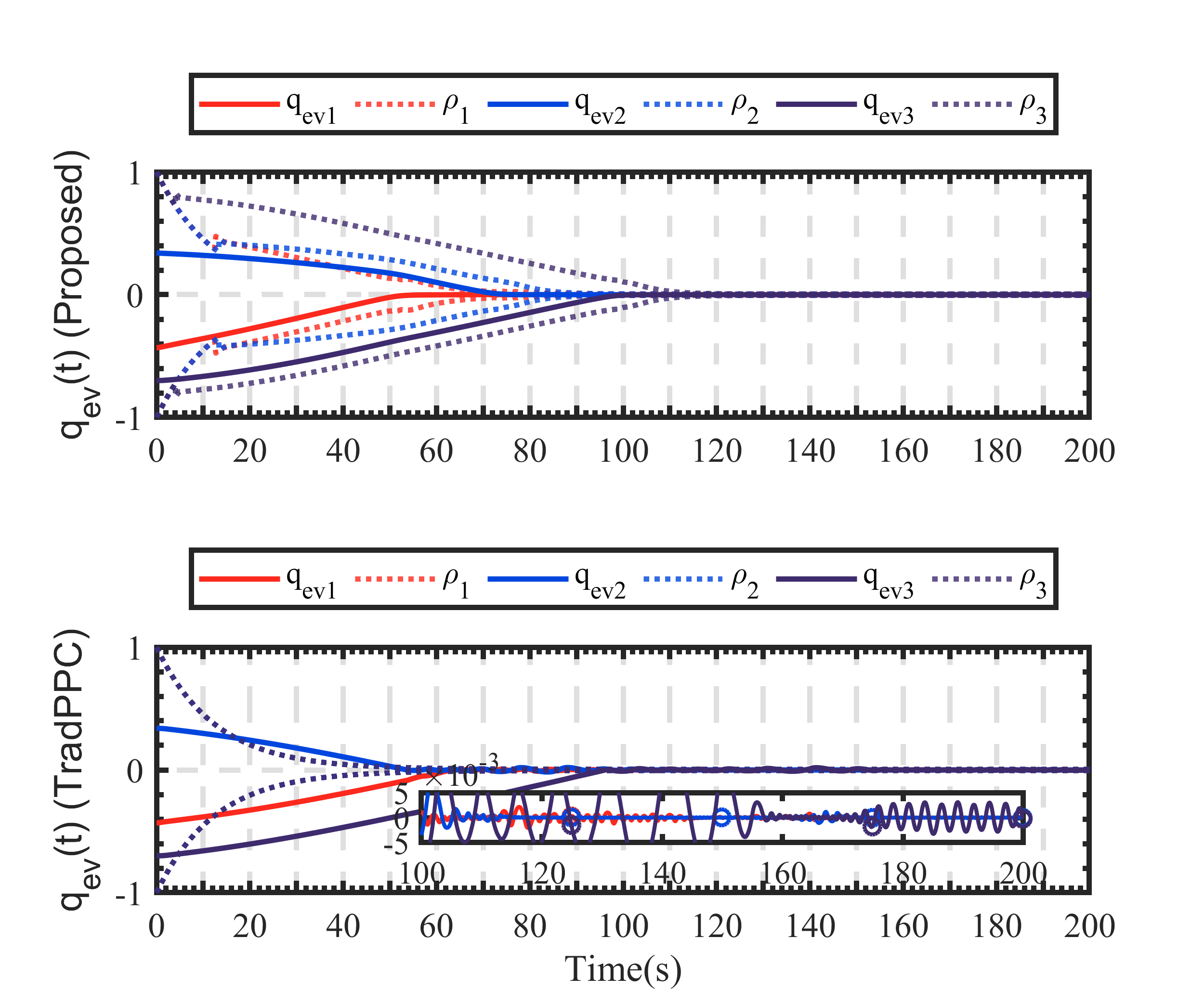}
	\caption{Time-Evolution of $\boldsymbol{q}_{ev}(t)$ with proposed ACPC Controller and \textit{TradPPC} Controller}       
	\label{CPCPPC}    
\end{figure}
Figure \ref{CPCPPC} shows the time-evolution of $q_{evi}(t)$ corresponds to the proposed controller and the \textit{TradPPC} benchmark controller. It can be observed that owing to the existing contradiction between angular velocity limitation and the preassigned performance envelope, the system is fail to keep inside the performance envelope, indicating that it is impossible for the benchmark \textit{TradPPC} controller to satisfy the performance envelope constraint and the angular velocity limitation simultaneously. By leveraging the detection and alleviation strategy, the proposed controller however, successfully handles the multiple constraint with a smooth control behavior. 

Notably, the \textit{TradPPC} scheme also appears a bad transient behavior when converges to the steady-state. This can be explained as the benchmark controller utilizes a Barrier Lyapunov function (e.g. $\ln\frac{1}{1-\varepsilon^{2}_{i}(t)}$) to realize the state constraint. Owing to the incompatibility problem, it is impossible for the state trajectory to remain within the performance envelope during the whole control process. However, as we previous stated in \cite{lei2023singularity}, once the state trajectory is not restricted by the performance envelope, the traditional PPC-based controller will suffer from the singularity problem, which will lead to a bad transient behavior. $\boldsymbol{q}_{ev}(t)$-trajectory will be strongly be battered by the performance envelope repeatedly, and thus leading to a chattering.
This validates our analysis in Subsection \ref{Motivation}.

	\section{CONCLUSION}
    This paper addresses the spacecraft attitude control problem while considering multiple constraints, such as performance requirements, limitations on angular velocity, and input saturation. Additionally, the time-varying parameter uncertainty is taken into account simultaneously. This paper proposes a Compatible Performance Control (CPC) scheme to address this issue completely.
   The CPC scheme integrates the conventional PPC framework with a contradiction detection mechanism based on a Zeroing Barrier Function (ZBF) and a strategy for modifying the bounded performance envelope. Simulations show that when the initially assigned nominal performance envelope is incompatible with the angular velocity constraint, a dynamic projection operator-governed modification generation dynamical system rapidly generates a bounded signal $\Delta\rho_{i}(t)$. This signal effectively expands the performance envelope, temporarily compromising on the performance envelope constraint and successfully resolving the contradiction problem. Once the contradiction no longer exists, the modification signal quickly vanishes, recovering the performance envelope to the nominal one. As a result, the proposed CPC scheme ensures the satisfaction of all constraints simultaneously during the whole control process.
   Furthermore, compared to a constant adaptive gain strategy, the designed time-varying gain adaptive strategy exhibits a smoother convergence behavior for the backstepping tracking error, i.e., the $\boldsymbol{z}_{2}$-subsystem. Consequently, it enhances the system's robustness against disturbances, improving disturbance rejection performance.

	\section{APPENDIX}
	\subsection{Proof of Main Stability Theorem \ref{T1}}\label{prooftheorem}
	\begin{proof}
		We first discuss on the stability of the output-system. In order to ensure that the performance envelope constraint is satisfied, i.e., $|\varepsilon_{i}(t)| < 1(i=1,2,3)$ holds for $\forall t\in\left[t_{0},+\infty\right)$, we introduce a Barrier Lyapunov function (BLF), denoted as $V_{B}$, to provide appropriate "constraining effect". The employed BLF function $V_{B}$ can be expressed as follows:
		\begin{equation}
			V_{B} =  \frac{k_{B}}{2}\sum_{i=1}^{3}\ln\frac{1}{1-\varepsilon^{2}_{i}(t)}(i=1,2,3)
		\end{equation}
		Taking the time-derivative of $V_{B}$, one can be obtained that:$\dot{V}_{B}  = k_{B}\sum_{i=1}^{3}\frac{\varepsilon_{i}\dot{\varepsilon_{i}}}{1-\varepsilon^{2}_{i}}(i=1,2,3)$. 
		Recalling the definition of $\boldsymbol{G}_{\varepsilon} = k_{B}\text{diag}\left(\frac{1}{1-\varepsilon^{2}_{i}}\right)(i=1,2,3)$ given in equation (\ref{vdesign}), it yields the compact form of $\dot{V}_{B}$. Accordingly, choosing a candidate Lyapunov function $V_{1}$ as:
		\begin{equation}
			V_{1} \triangleq V_{B} + \frac{1}{2}\boldsymbol{\omega}^{\text{T}}_{v}\boldsymbol{\omega}_{v}+\frac{K_{q}}{2}\boldsymbol{q}^{\text{T}}_{ev}\boldsymbol{q}_{ev}
		\end{equation}
		
		Therefore, taking the time-derivative of $V_{1}$ and combined with the result given by equation (\ref{closeloop2}) and (\ref{vdesign}), we have:
		\begin{equation}\label{dV1}
			\begin{aligned}
				\dot{V}_{1} 
				&= -C_{d}C_{f}\|\boldsymbol{\omega}_{v}\|_{2}^{2} \\
				&\quad+ \left(\boldsymbol{\varepsilon}^{\text{T}}\boldsymbol{G}_{\varepsilon}\boldsymbol{\Psi}\boldsymbol{\varGamma}_{e}+K_{q}\boldsymbol{q}^{\text{T}}_{ev}\boldsymbol{\varGamma}_{e}\right)\boldsymbol{z}_{2} + K_{\xi}\boldsymbol{\omega}^{\text{T}}_{v}\boldsymbol{\xi}\\
				&\le
				-\left(C_{d}C_{f}-\frac{K_{\xi}}{2}\right)\|\boldsymbol{\omega}_{v}\|^{2}_{2}+\boldsymbol{z}^{\text{T}}_{2}\boldsymbol{P}_{\varepsilon} + \frac{K_{\xi}}{2}\|\boldsymbol{\xi}\|_{2}^{2}
			\end{aligned}
		\end{equation}
		Note that $\boldsymbol{P}_{\varepsilon}$ that previously defined below equation (\ref{uc}), $\left(\boldsymbol{\varepsilon}^{\text{T}}\boldsymbol{G}_{\varepsilon}\boldsymbol{\Psi}\boldsymbol{\varGamma}_{e}+K_{q}\boldsymbol{q}^{\text{T}}_{ev}\boldsymbol{\varGamma}_{e}\right)\boldsymbol{z}_{2}$ can be expressed as $\boldsymbol{z}^{\text{T}}_{2}\boldsymbol{P}_{\varepsilon}$. By applying the Peter-Paul's inequality \cite{jarchow2012locally}, it can be further derived that:
		\begin{equation}\label{dotV1}
			\begin{aligned}
				\dot{V}_{1} &\le -\left(C_{d}C_{f}-\frac{K_{\xi}}{2}\right)\|\boldsymbol{\omega}_{v}\|_{2}^{2} + \frac{K_{\xi}}{2}\|\boldsymbol{\xi}\|_{2}^{2} \\
				&\quad + \frac{b_{1}}{2}+\frac{1}{2b_{1}}\|\boldsymbol{z}_{2}\|_{2}^{2}\|\boldsymbol{P}_{\varepsilon}\|_{2}^{2}
			\end{aligned}
		\end{equation}
		where $b_{1} > 0$ is a positive constant. We then defining two Lyapunov functions for $\boldsymbol{\xi}$-dynamics and $\Delta\boldsymbol{\rho}$-dynamics as:
		\begin{equation}
			\begin{aligned}
				V_{\xi} \triangleq \frac{1}{2}\boldsymbol{\xi}^{\text{T}}\boldsymbol{\xi};V_{\rho}\triangleq\frac{1}{2}\Delta\boldsymbol{\rho}^{\text{T}}\Delta\boldsymbol{\rho}
			\end{aligned}
		\end{equation}
		Taking the time-derivative of $V_{\xi}$ and combined with the design of $\boldsymbol{\xi}$ depicted in equation (\ref{xi}), one can be obtained that:
		\begin{equation}\label{Vxi}
			\begin{aligned}
				\dot{V}_{\xi} &= -p_{1}\boldsymbol{\xi}^{\text{T}}\boldsymbol{\xi} - p_{\xi}\|\Delta\boldsymbol{\varsigma}\|^{2} - p_{\xi}\|\boldsymbol{w}\|_{2}^{2} + p_{2}\boldsymbol{\xi}^{\text{T}}\left(\Delta\boldsymbol{\varsigma}+\boldsymbol{w}\right)-p_{c}\\
				&\le -\left(p_{1}-p_{2}\right)\|\boldsymbol{\xi}\|_{2}^{2} - \left(p_{\xi}-\frac{p_{2}}{2}\right)\left(\|\Delta\boldsymbol{\varsigma}\|_{2}^{2}+\|\boldsymbol{w}\|^{2}_{2}\right)-p_{c}
			\end{aligned}
		\end{equation}
		Meanwhile, taking the time-derivative of $V_{\rho}$ and combined with the design in equation (\ref{modified}), one can be yielded that:
		\begin{equation}\label{Vrho}
			\begin{aligned}
				\dot{V}_{\rho}&=\Delta\boldsymbol{\rho}^{\text{T}}\boldsymbol{G}_{\rho}\left[-C_{\rho}\Delta\boldsymbol{\rho}+C_{v}\Delta\boldsymbol{\varsigma}\right]-\Delta\boldsymbol{\rho}^{\text{T}}\boldsymbol{w}\\
				&\le -C_{\rho}G_{\rho\min}\|\Delta\boldsymbol{\rho}\|^{2}_{2}+C_{v}\|\Delta\boldsymbol{\rho}\|_{2}\|\Delta\boldsymbol{\varsigma}\|_{2}+\|\Delta\boldsymbol{\rho}\|_{2}\|\boldsymbol{w}\|_{2}\\
				&\le-\left(C_{\rho}G_{\rho\min}-\frac{C_{v}+1}{2}\right)\|\Delta\boldsymbol{\rho}\|^{2}_{2}+\frac{C_{v}}{2}\|\Delta\boldsymbol{\varsigma}\|^{2}_{2}+\frac{1}{2}\|\boldsymbol{w}\|^{2}_{2}
			\end{aligned}
		\end{equation}
		where $G_{\rho\min}$ stands for the minimum eigenvalue of $\boldsymbol{G}_{\varepsilon}$.
		Accordingly, combining these derived results yields:
		\begin{equation}\label{dotV1all1}
			\begin{aligned}
				&\dot{V}_{1} +\dot{V}_{\xi} +\dot{V}_{\rho}\\
				\le & -\left(C_{d}C_{f}-\frac{K_{\xi}}{2}\right)\|\boldsymbol{\omega}_{v}\|_{2}^{2}-\left(2p_{1}-2p_{2}-K_{\xi}\right)\frac{1}{2}\|\boldsymbol{\xi}\|^{2}_{2}\\
				&-\left(2C_{\rho}G_{\rho\min}-C_{v}-1\right)\frac{1}{2}\|\Delta\boldsymbol{\rho}\|_{2}^{2}\\
				&-\left(p_{\xi}-\frac{p_{2}+C_{v}}{2}\right)\|\Delta\boldsymbol{\varsigma}\|_{2}^{2}-\left(p_{\xi}-\frac{p_{2}+1}{2}\right)\|\boldsymbol{w}\|^{2}_{2}\\
				&+\frac{b_{1}}{2}+\frac{1}{2b_{1}}\|\boldsymbol{z}_{2}\|_{2}^{2}\|\boldsymbol{P}_{\varepsilon}\|_{2}^{2}-p_{c}
			\end{aligned}
		\end{equation}
		Notably, for each element of the diagonal matrix $\boldsymbol{G}_{\rho}$, its minimum value is subjected by $g_{i}(t) \ge 0$. Meanwhile, according to equation (\ref{rhomax}), we have $\Delta\rho_{i}(t) \le |q_{evi}(t)| + \sigma_{\rho} - \rho_{ni}(t) < |q_{evi}(t)|+\sigma_{\rho}$. Therefore, it can be derived that:
		\begin{equation}\label{drho}
			\sum_{j=1}^{3}\frac{1}{2}(C_{v}+1)\Delta\rho^{2}_{i} \le\frac{3(C_{v}+1)}{2}(1+\sigma_{\rho})^{2}\triangleq L_{\rho}
		\end{equation}
	     the right hand side of equation (\ref{drho}) is defined as $L_{\rho}$, which is a constant that determined by parameter selection.
		We further define $G_{\xi}\triangleq 2p_{1}-2p_{2}-K_{\xi}$, $G_{\varsigma}\triangleq p_{\xi}-\frac{p_{2}+C_{v}}{2}$ and $G_{w}\triangleq p_{\xi}-\frac{p_{2}+1}{2}$ for brevity. Meanwhile, choosing parameter as $p_{c} = L_{\rho}$, equation (\ref{dotV1all1}) can be rearranged as follows:
		\begin{equation}\label{dotV1all}
			\begin{aligned}
				&\dot{V}_{1} + \dot{V}_{\xi}+\dot{V}_{\rho}\\
				\le& -\left(C_{d}C_{f}-\frac{K_{\xi}}{2}\right)\|\boldsymbol{\omega}_{v}\|_{2}^{2}-G_{\varsigma}\|\Delta\boldsymbol{\varsigma}\|_{2}^{2}-G_{w}\|\boldsymbol{w}\|^{2}_{2}
				\\
				&+\frac{b_{1}}{2}+\frac{1}{2b_{1}}\|\boldsymbol{z}_{2}\|_{2}^{2}\|\boldsymbol{P}_{\varepsilon}\|_{2}^{2}-G_{\xi}\|\boldsymbol{\xi}\|^{2}_{2}
			\end{aligned}
		\end{equation}
		
		Subsequently, we further choosing the candidate Lyapunov function for $\boldsymbol{z}_{2}$-system as: $V_{2} = \frac{1}{2}\boldsymbol{z}^{\text{T}}_{2}\boldsymbol{J}_{0}\boldsymbol{z}_{2}$.
		Considering the time-derivative of $V_{2}$, substituting the design of $\boldsymbol{u}_{c}$ depicted in equation (\ref{uc}) into is, one can be obtained that:
		\begin{equation}
			\begin{aligned}
				\dot{V}_{2} &= -K_{2}\boldsymbol{z}^{\text{T}}_{2}\boldsymbol{J}_{0}\boldsymbol{z}_{2}+K_{\delta}\boldsymbol{z}^{\text{T}}_{2}\boldsymbol{\delta}+K_{\gamma}\boldsymbol{z}^{\text{T}}_{2}\boldsymbol{\gamma}_{\varepsilon}\\
				&\quad +\boldsymbol{z}^{\text{T}}_{2}\boldsymbol{W}\left(\boldsymbol{\Theta} - \hat{\boldsymbol{\Theta}}\right) + \boldsymbol{z}^{\text{T}}_{2}\Delta\boldsymbol{u}
			\end{aligned}
		\end{equation}
		To facilitate the following analysis, we further defining an estimation error variable $\tilde{\boldsymbol{\Theta}}\in\mathbb{R}^{9}$ as $\tilde{\boldsymbol{\Theta}} \triangleq \hat{\boldsymbol{\Theta}} - \boldsymbol{\Theta}$. Notably, $\frac{1}{2}\lambda_{J0\min}\|\boldsymbol{z}_{2}\|_{2}^{2}\le V_{2}$ holds, thus it can be yielded that $\|\boldsymbol{z}_{2}\|_{2}^{2}\le \frac{2V_{2}}{\lambda_{J0\min}}$. By applying the Peter-Paul inequality, one can be further obtained that:
		\begin{equation}\label{dotV2}
			\begin{aligned}
				\dot{V}_{2} &\le -K_{2}\boldsymbol{z}^{\text{T}}_{2}\boldsymbol{J}_{0}\boldsymbol{z}_{2} + \frac{K_{\delta}a_{1}}{2}\|\boldsymbol{z}_{2}\|_{2}^{2} + \frac{K_{\delta}}{2a_{1}}\|\boldsymbol{\delta}\|^{2}_{2}\\
				&\quad + \frac{K_{\gamma}a_{2}}{2}\|\boldsymbol{z}_{2}\|^{2}_{2}+\frac{K_{\gamma}}{2a_{2}}\|\boldsymbol{\gamma}_{\varepsilon}\|^{2}_{2} \\
				&\quad + \frac{a_{3}}{2}\|\boldsymbol{z}_{2}\|_{2}^{2} +\frac{1}{2a_{3}}\|\Delta\boldsymbol{u}\|_{2}^{2} - \boldsymbol{z}^{\text{T}}_{2}\boldsymbol{W}\tilde{\boldsymbol{\Theta}}\\
				&\le\ -\left[2K_{2}-\frac{K_{\delta}a_{1}+K_{\gamma}a_{2} + a_{3}}{\lambda_{J0\min}}\right]V_{2}-\boldsymbol{z}^{\text{T}}_{2}\boldsymbol{W}\tilde{\boldsymbol{\Theta}}\\
				&\quad + \frac{K_{\delta}}{2a_{1}}\|\boldsymbol{\delta}\|_{2}^{2} + \frac{K_{\gamma}}{2a_{2}}\|\boldsymbol{\gamma}_{\varepsilon}\|_{2}^{2}+\frac{1}{2a_{3}}\|\Delta\boldsymbol{u}\|_{2}^{2}
			\end{aligned}
		\end{equation}
		where $a_{1}, a_{2}, a_{3} > 0$ are positive constants. 
		
		Next, considering a candidate Lyapunov function for $\boldsymbol{\gamma}_{\varepsilon}$-system and $\boldsymbol{\delta}$-system as:
		\begin{equation}
			V_{\gamma}\triangleq\frac{1}{2}\boldsymbol{\gamma}^{\text{T}}_{\varepsilon}\boldsymbol{\gamma}_{\varepsilon};V_{\delta}\triangleq\frac{1}{2}\boldsymbol{\delta}^{\text{T}}\boldsymbol{\delta}
		\end{equation}
		Taking the time-derivative of $V_{\gamma}$ and combined with the design in equation (\ref{gamma}), by applying the Young's inequality, it can be yielded that:
		\begin{equation}\label{dotVgamma}
			\begin{aligned}
				\dot{V}_{\gamma} & = -g_{1}\|\boldsymbol{\gamma}_{\varepsilon}\|_{2}^{2}-g_{\gamma}\|\boldsymbol{z}_{2}\|_{2}^{2}\|\boldsymbol{P}_{\varepsilon}\|_{2}^{2}+g_{2}\boldsymbol{\gamma}^{\text{T}}_{\varepsilon}\|\boldsymbol{P}_{\varepsilon}\|\boldsymbol{z}_{2}-g_{c}\\
				&\le -\left(g_{1}-\frac{g_{2}}{2}\right)\|\boldsymbol{\gamma}_{\varepsilon}\|_{2}^{2} - \left(g_{\gamma}-\frac{g_{2}}{2}\right)\|\boldsymbol{P}_{\varepsilon}\|_{2}^{2}\|\boldsymbol{z}_{2}\|_{2}^{2}-g_{c}
			\end{aligned}
		\end{equation}
		Similarly, taking the time-derivative of $V_{\delta}$, it can be derived that:
		\begin{equation}\label{dotVdelta}
			\begin{aligned}
				\dot{V}_{\delta} &\le -\left(n_{1}-\frac{n_{2}}{2}\right)\|\boldsymbol{\delta}\|_{2}^{2}-\left(n_{\delta}-\frac{n_{2}}{2}\right)\|\Delta\boldsymbol{u}\|_{2}^{2}
			\end{aligned}
		\end{equation}
		Therefore, combining the result in equation (\ref{dotV2})(\ref{dotVgamma})(\ref{dotVdelta}), it can be yielded that:
		\begin{equation}
			\begin{aligned}
				&\dot{V}_{2} + \dot{V}_{\delta} + \dot{V}_{\gamma}\\
				\le& -\left[2K_{2}-\frac{a_{1}K_{\delta}+a_{2}K_{\gamma}+a_{3}}{\lambda_{J0\min}}\right]V_{2}-\boldsymbol{z}^{\text{T}}_{2}\boldsymbol{W}\tilde{\boldsymbol{\Theta}}\\
				&-\left[2g_{1}-g_{2}-\frac{K_{\gamma}}{a_{2}}\right]V_{\gamma} - \left[2n_{1}-n_{2}-\frac{K_{\delta}}{a_{1}}\right]V_{\delta}-g_{c}\\
				&-\left(g_{\gamma}-\frac{g_{2}}{2}\right)\|\boldsymbol{z}_{2}\|_{2}^{2}\|\boldsymbol{P}_{\varepsilon}\|_{2}^{2}-\left[n_{\delta}-\frac{n_{2}}{2}-\frac{1}{2a_{3}}\right]\|\Delta\boldsymbol{u}\|_{2}^{2}\\
			\end{aligned}
		\end{equation} 
		For the brevity, we define $G_{2}\triangleq 2K_{2}-\frac{a_{1}K_{\delta}+a_{2}K_{\gamma}+a_{3}}{\lambda_{J0\min}}$, $G_{\gamma}\triangleq2g_{1}-g_{2}-\frac{K_{\gamma}}{a_{2}}$, $G_{\delta} \triangleq 2n_{1}-n_{2}-\frac{K_{\delta}}{a_{1}}$, $G_{P}\triangleq g_{\gamma}-\frac{g_{2}}{2}$ and $G_{u}\triangleq n_{\delta}-\frac{n_{2}}{2}-\frac{1}{2a_{3}}$. Thereby, the above equation can be simplified as:
		\begin{equation}\label{dotV2deltagamma}
			\begin{aligned}
				&\dot{V}_{2}+\dot{V}_{\delta}+\dot{V}_{\gamma}\\
				\le& -G_{2}V_{2}-G_{\gamma}V_{\gamma}-G_{\delta}V_{\delta}-G_{P}\|\boldsymbol{z}_{2}\|_{2}^{2}\|\boldsymbol{P}_{\varepsilon}\|_{2}^{2}\\
				&-G_{u}\|\Delta\boldsymbol{u}\|_{2}^{2}
				-\boldsymbol{z}^{\text{T}}_{2}\boldsymbol{W}\tilde{\boldsymbol{\Theta}}-g_{c}
			\end{aligned}
		\end{equation}
		Subsequently, choosing the candidate Lyapunov function of parameter estimation error variable $\tilde{\boldsymbol{\Theta}}$ as $V_{\Theta}$, expressed as: 
		\begin{equation}
			V_{\Theta} \triangleq \frac{1}{2\zeta(t)}\tilde{\boldsymbol{\Theta}}^{\text{T}}\tilde{\boldsymbol{\Theta}}
		\end{equation}
		Taking the time-derivative of $V_{\Theta}$, it can be obtained that:
		\begin{equation}
			\begin{aligned}
				\dot{V}_{\Theta}
				&= -\frac{1}{2\zeta^{2}(t)}\dot{\zeta}(t)\tilde{\boldsymbol{\Theta}}^{\text{T}}\tilde{\boldsymbol{\Theta}}+\frac{1}{\zeta(t)}\tilde{\boldsymbol{\Theta}}^{\text{T}}\dot{\hat{\boldsymbol{\Theta}}} - \frac{1}{\zeta(t)}\tilde{\boldsymbol{\Theta}}^{\text{T}}\dot{\boldsymbol{\Theta}}
			\end{aligned}
		\end{equation}
		For the convenience of expression, let $\Theta_{r}(t)$ defined as $\Theta_{r}(t) \triangleq \frac{1}{\zeta(t)}\|\tilde{\boldsymbol{\Theta}}\|_{2}\|\dot{\boldsymbol{\Theta}}\|_{2}$.
		Substituting the design of $\dot{\zeta}(t)$ stated in equation (\ref{zetacompact}) and the design of $\dot{\hat{\boldsymbol{\Theta}}}$ specified by equation (\ref{theta}) into the expression of $\dot{V}_{\Theta}$, one can be yielded that:
		\begin{equation}\label{dV3}
			\begin{aligned}
				\dot{V}_{\Theta} 
				& \le -\lambda_{\zeta}\mu(t)\left[\frac{1}{\zeta(t)}-K_{\zeta}\chi(t)\right]\|\tilde{\boldsymbol{\Theta}}\|^{2}_{2}(t) + \Theta_{r}\\
				&\quad + \frac{1}{\zeta(t)}\tilde{\boldsymbol{\Theta}}^{\text{T}}\text{Proj}\left(\hat{\boldsymbol{\Theta}},\zeta(t)\boldsymbol{W}^{\text{T}}\boldsymbol{z}_{2},f(\hat{\boldsymbol{\Theta}})\right)\\
			\end{aligned}
		\end{equation}
		Combining the result in equation (\ref{dotV2deltagamma}) and (\ref{dV3}), one can be further obtained that:
		\begin{equation}\label{dotV2all1}
			\begin{aligned}
				&\dot{V}_{2}+\dot{V}_{\Theta}+\dot{V}_{\delta}+\dot{V}_{\gamma}\\
				\le& -G_{2}V_{2} - G_{\delta}V_{\delta}-G_{\gamma}V_{\gamma}-G_{P}\|\boldsymbol{z}_{2}\|_{2}^{2}\|\boldsymbol{P}_{\varepsilon}\|_{2}^{2}-G_{u}\|\Delta\boldsymbol{u}\|_{2}^{2}\\
				&+\frac{1}{\zeta(t)}\tilde{\boldsymbol{\Theta}}^{\text{T}}\left[\text{Proj}\left(\hat{\boldsymbol{\Theta}},\zeta(t)\boldsymbol{W}^{\text{T}}\boldsymbol{z}_{2},f(\tilde{\boldsymbol{\Theta}})\right)-\zeta(t)\boldsymbol{W}^{\text{T}}\boldsymbol{z}_{2}\right]\\
				&-\lambda_{\zeta}\mu(t)\left[\frac{1}{\zeta(t)}-K_{\zeta}\chi(t)\right]\|\tilde{\boldsymbol{\Theta}}\|_{2}^{2}+\Theta_{r}(t)-g_{c}
			\end{aligned}
		\end{equation}
		
		We first discuss on the symbol of the middle term, defined as $L_{\Theta}$ for writing convenience, expressed as:
		\begin{equation}
			\begin{aligned}
				L_{\Theta}\triangleq\frac{1}{\zeta(t)}\tilde{\boldsymbol{\Theta}}^{\text{T}}\left[\text{Proj}\left(\hat{\boldsymbol{\Theta}},\zeta(t)\boldsymbol{W}^{\text{T}}\boldsymbol{z}_{2},f(\hat{\boldsymbol{\Theta}})\right)-\zeta(t)\boldsymbol{W}^{\text{T}}\boldsymbol{z}_{2}\right]
			\end{aligned}
		\end{equation}
		Considering the expression of $\text{Proj}(\hat{\boldsymbol{\Theta}},\zeta(t)\boldsymbol{W}^{\text{T}}\boldsymbol{z}_{2},f(\hat{\boldsymbol{\Theta}}))$, for the second condition in equation (\ref{theta}), i.e., $\dot{\hat{\boldsymbol{\Theta}}} = \zeta(t)\boldsymbol{W}^{\text{T}}\boldsymbol{z}_{2}$, it can be obtained that $L_{\Theta} = 0$ holds; For the first condition in equation (\ref{theta}), it can be yielded that:
		\begin{equation}
			L_{\Theta} \triangleq -\frac{1}{\zeta(t)}\tilde{\boldsymbol{\Theta}}^{\text{T}}\frac{\boldsymbol{\nabla}_{\Theta}\boldsymbol{\nabla}^{\text{T}}_{\Theta} \cdot f(\hat{\boldsymbol{\Theta}})}{\|\boldsymbol{\nabla}_{\Theta}\|_{2}^{2}}\zeta(t)\boldsymbol{W}^{\text{T}}\boldsymbol{z}_{2}
		\end{equation}
		According to Lemma \ref{LemmaPartial}, since the true parameter is an interior point such that $f(\boldsymbol{\Theta}) \le 0$ holds, we have $\tilde{\boldsymbol{\Theta}}^{\text{T}}\boldsymbol{\nabla}_{\Theta} \ge 0$. Meanwhile, according to the condition for the first circumstance in equation (\ref{theta}), we have $f(\hat{\boldsymbol{\Theta}}) > 0$ and $\boldsymbol{\nabla}^{\text{T}}_{\Theta}\zeta(t)\boldsymbol{W}^{\text{T}}\boldsymbol{z}_{2} > 0$. According to the analysis in Corollary \ref{coroZeta}, it can be obtained that $\frac{1}{\zeta(t)} > 0$ holds, thereby this concludes the final result that $L_{\Theta} \le  0$ will be satisfied for the first condition. Consequently, it can be obtained that $L_{\Theta}\le 0$ is always satisfied.
		Substituting this result into equation (\ref{dotV2all1}), it can be further obtained that:
		\begin{equation}\label{dotV2all}
			\begin{aligned}
				&\dot{V}_{2}+\dot{V}_{\Theta}+\dot{V}_{\delta}+\dot{V}_{\gamma}\\
				\le& -G_{2}V_{2} - G_{\delta}V_{\delta}-G_{\gamma}V_{\gamma}\\
				&-G_{P}\|\boldsymbol{z}_{2}\|_{2}^{2}\|\boldsymbol{P}_{\varepsilon}\|_{2}^{2}-G_{u}\|\Delta\boldsymbol{u}\|_{2}^{2}\\
				&-\lambda_{\zeta}\mu(t)\left[\frac{1}{\zeta(t)}-K_{\zeta}\chi(t)\right]\|\tilde{\boldsymbol{\Theta}}\|_{2}^{2}+\Theta_{r}(t)-g_{c}
			\end{aligned}
		\end{equation}
		
		Defining a lumped Lyapunov function $V$ as the sum of all defined Lyapunov functions, expressed as: $V\triangleq V_{1}+V_{\xi}+V_{\rho}+V_{2}+V_{\Theta}+V_{\delta}+V_{\gamma}$. Choosing parameter as $g_{c} = \frac{b_{1}}{2}$, by combining the result given by equation (\ref{dotV1all}) and (\ref{dotV2all}), it yields the following relationship:
		\begin{equation}\label{dotVfinal1}
			\begin{aligned}
				\dot{V}
				&\le -\left(C_{d}C_{f}-\frac{K_{\xi}}{2}\right)\|\boldsymbol{\omega}_{v}\|_{2}^{2}-G_{\xi}V_{\xi}-G_{w}\|\boldsymbol{w}\|^{2}_{2}\\
				&\quad -G_{\varsigma}\|\Delta\boldsymbol{\varsigma}\|_{2}^{2}
				-G_{2}V_{2} - G_{\delta}V_{\delta}-G_{\gamma}V_{\gamma}\\
				&\quad-\left(G_{P}-\frac{1}{2b_{1}}\right)\|\boldsymbol{z}_{2}\|_{2}^{2}\|\boldsymbol{P}_{\varepsilon}\|_{2}^{2}-G_{u}\|\Delta\boldsymbol{u}\|_{2}^{2}\\
				&\quad-\lambda_{\zeta}\mu(t)\left[\frac{1}{\zeta(t)}-K_{\zeta}\chi(t)\right]\|\tilde{\boldsymbol{\Theta}}\|_{2}^{2}+\Theta_{r}(t)
			\end{aligned}
		\end{equation}
		Notably, in order to facilitate the stability analysis, the parameter selecting should ensure that $C_{d}C_{f}-\frac{K_{\xi}}{2}$, $G_{2}$, $G_{P}-\frac{1}{2b_{1}}$, $G_{\xi}$, $G_{\varsigma}$, $G_{\delta}$, $G_{\gamma}$, $G_{w}$ and $G_{u}$ are all positive parameters.
		
		From the expression of $\dot{V}$ stated in equation (\ref{dotVfinal1}), the final conclusion of the stability analysis is discussed for two circumstances $\mu(t) = 0$ and $\mu(t) \in\left(0,1\right]$ separately.
		
		\textbf{1.} For the circumstance that $\mu(t) = 0$ holds, this refers to the condition that $f(\zeta(t)) = 1$. Considering the expression of $\Theta_{r}(t)$:
		\begin{equation}
			\begin{aligned}
				\Theta_{r}(t) = \frac{1}{\zeta(t)}\|\tilde{\boldsymbol{\Theta}}\|_{2}\|\dot{\boldsymbol{\Theta}}\|_{2}
			\end{aligned}
		\end{equation}
		According to Corollary \ref{coroZeta}, we have $\frac{1}{\zeta(t)} \le \frac{1}{\zeta(t_{0})} + K_{\zeta}$. On the other hand, owing to the fact that we have designed a projection operator-based adaptive law for the updating of $\hat{\boldsymbol{\Theta}}$, while the unknown parameter is assumed to be bounded according to Assumption \ref{BoundedUnknown}, thereby this guarantees that there exists a maxima of $\|\tilde{\boldsymbol{\Theta}}\|_{2}$ during the whole control process, and we further denote it as $\tilde{\Theta}_{\max}$. Meanwhile, the time-derivative of the varying parameter is assumed to satisfy $\|\dot{\boldsymbol{\Theta}}\|_{2} \le \Theta_{d\max}$. Thus, we have $\Theta_{r}(t) \le \left(\frac{1}{\zeta(t_{0})}+K_{\zeta}\right)\tilde{\Theta}_{\max}\Theta_{d\max}$, and we further denote the right term as $\Theta_{r\max} \triangleq \left(\frac{1}{\zeta(t_{0})}+K_{\zeta}\right)\tilde{\Theta}_{\max}\Theta_{d\max}$.
		
		Defining a lumped Lyapunov term such that the right hand side of equation (\ref{dotVfinal1}) can be expressed as $-A_{v}+\Theta_{r\max}$, i.e., $\dot{V} \le -A_{v}+\Theta_{r\max}$ holds, then it can be easily obtained that $\dot{V}\le 0$ holds for $A_{v} \ge \Theta_{r\max}$. Since we have provided in Corollary \ref{coroChi} and \ref{coroZeta} that $\zeta(t) < \zeta_{\max}$, $\chi(t)\in\left[0,1\right]$ will be satisfied for arbitrary $\forall t\in\left[t_{0},+\infty\right)$ without providing any additional condition, this ensures the boundedness of all closed-loop signals, which further completes the proof of the closed-loop system's boundedness.
				
%		Notably, for the special case that the unknown parameter are not time-varying, i.e. $\dot{\boldsymbol{\Theta}} = \boldsymbol{0}$, then $\Theta
%		_{r}(t) = 0$ can be derived. Therefore, the negative of $\dot{V}$ is strict, and the boundedness conclusion for $\mu(t) = 0$ can be strengthened to be "the system will asymptotically converged".
		
		\textbf{2.} Next, we considering the circumstance that $\mu(t) \in\left(0,1\right]$. It can be observed that for $\left(\frac{1}{\zeta(t)}-K_{\zeta}\chi(t)\right)>0$, the term $-\lambda_{\zeta}\mu(t)\left(\frac{1}{\zeta(t)}-K_{\zeta}\chi(t)\right) < 0$ holds, therefore, compared with the first discussed circumstance that $\mu(t) = 0$, $-\lambda_{\zeta}\mu(t)\left(\frac{1}{\zeta(t)}-K_{\zeta}\chi(t)\right) < 0$ provides an additional negative contribution to $\dot{V}$. Following a similar analysis, this proofs the boundedness of the closed-loop system.
		
		For $\frac{1}{\zeta(t)} - K_{\zeta}\chi(t) < 0$, it is obvious that the boundedness of the closed-loop system can be "conservatively" derived as:
		\begin{equation}
			\begin{aligned}
				&-\lambda_{\zeta}\mu(t)\left(\frac{1}{\zeta(t)}-K_{\zeta}\chi(t)\right)\|\tilde{\boldsymbol{\Theta}}\|_{2}^{2}+\Theta_{r}(t)\\ \le&\lambda_{\zeta}\left(K_{\zeta}-\frac{1}{\zeta_{\max}}\right)\tilde\Theta^{2}_{\max}+\left(\frac{1}{\zeta(t_{0})}+K_{\zeta}\right)\tilde{\Theta}_{\max}\Theta_{d\max}\\
			\end{aligned}
		\end{equation}
		Therefore, following a similar analysis as in the first circumstance with $\mu(t) = 0$, it concludes the boundedness of the closed-loop system.
		Additionally, 
		 it can be observed that $\frac{1}{\zeta(t)} - K_{\zeta}\chi(t) < 0$ can be equivalently expressed as $\zeta(t) > \frac{1}{K_{\zeta}\chi(t)}$, indicating that this circumstance occurs when $\zeta(t)$ is sufficiently large. On the one hand, note that under this circumstance, $\mathcal{Z}(t)$ becomes negative and such $\dot{\zeta}(t) < 0$. Accordingly, the value of $\zeta(t)$ will immediately decreasing. On the other hand, since $\|\boldsymbol{W}\|$ decreases as closed-loop system converges, then $\chi(t)$ becoming smaller such that $\frac{1}{K_{\zeta}\chi(t)}$ rapidly increases. Therefore, this also making $K_{\zeta}\chi(t)$ smaller and finally leading to $\frac{1}{\zeta(t)} > K_{\zeta}\chi(t)$. Therefore, these analysis indicates that the system will finally turns to satisfy $\left(\frac{1}{\zeta(t)}-K_{\zeta}\chi(t)\right)>0$ as the system converges, which provides additional negative contribution for $\dot{V}$.
		
		Correspondingly, this completes the proof of the system's boundedness in Theorem \ref{T1}.

	\end{proof}
	
	\subsection{Proof of Corollary \ref{coroChi}}\label{ProofChi}
	\begin{proof}
		Considering the expression of $\dot{\chi}(t)$ given in equation (\ref{chi}), we have:
		\begin{equation}
			\dot{\chi}(t) = -\lambda_{\chi}\chi(t) + \lambda_{\chi}\frac{\|\boldsymbol{W}\|^{2}_{2}}{1+\|\boldsymbol{W}\|^{2}_{2}}
		\end{equation}
		Let the initial condition of $\chi(t)$ as $\chi(t_{0})\in\left[0,1\right]$, integrating on both sides of the above equation, one can be yielded that:
		\begin{equation}\label{chitime}
			\begin{aligned}
				\chi(t) &= e^{-\lambda_{\chi}(t-t_{0})}\chi(t_{0}) + \lambda_{\chi}\int_{t_{0}}^{t}e^{-\lambda_{\chi}(t-s)}\frac{\|\boldsymbol{W}(s)\|_{2}^{2}}{1+\|\boldsymbol{W}(s)\|^{2}_{2}}ds 
			\end{aligned}
		\end{equation}
		therefore, owing to the fact that $\chi(t_{0})$ have been chosen to be positive, and each term in this integration is non-negative, it can be obtained that $\chi(t) \ge 0$ holds for $\forall t\in\left[t_{0},+\infty\right)$. 
		
		Meanwhile, it can be observed that $\frac{\|\boldsymbol{W}\|^{2}_{2}}{1+\|\boldsymbol{W}\|^{2}_{2}} < 1$ will be always hold, thus we have:
		\begin{equation}\label{chixiaoyuyi}
			\begin{aligned}
				\chi(t) &< e^{-\lambda_{\chi}(t-t_{0})}\chi(t_{0}) + \lambda_{\chi}\int_{t_{0}}^{t}e^{-\lambda_{\chi}(t-s)}ds\\
				&=1-e^{-\lambda_{\chi}(t-t_{0})}\left(1-\chi(t_{0})\right)
			\end{aligned}
		\end{equation}
		Owing to the fact that we have chosen $\chi(t_{0})\in\left[0,1\right]$, it yields $\chi(t)<1$ holds for $\forall t\in\left[t_{0},+\infty\right)$.

		this completes the proof of Corollary \ref{coroChi}.
	\end{proof}

	\subsection{Proof of Corollary \ref{coroZeta}}\label{ProofZeta}
	\begin{proof}
		We first discuss on the factor $\mu(t)$ in the $\zeta(t)$-dynamics. Considering the evaluation function $f(\zeta(t))$, taking its time-derivative and combining with different conditions of $\zeta(t)$-dynamics, it can be yielded that:
		\begin{equation}\label{dotfzeta}
			\begin{aligned}
				\dot{f}(\zeta(t)) &= \lambda_{\zeta}\nabla_{\zeta}\text{Proj}\left[\zeta(t),\mathcal{Z}(t),f(\zeta(t))\right]\\
				&=
				\begin{cases}
					\dot{f}(\zeta(t)) >0,f(\zeta(t))\in\left(0,1\right)\text{and}\nabla_{\zeta}\mathcal{Z}(t)>0\\
					\dot{f}(\zeta(t))<0,\nabla_{\zeta}\mathcal{Z}(t) < 0\\
					\dot{f}(\zeta(t)) = 0,f(\zeta(t)) = 1\text{and}\nabla_{\zeta}\mathcal{Z}(t) >0
				\end{cases}
			\end{aligned}
		\end{equation}
		Accordingly, it can be obtained from equation (\ref{dotfzeta}) and Remark \ref{mucharas} that $\mu(t)\in\left[0,1\right]$ holds for $\forall t\in\left[t_{0},+\infty\right)$.
		
		Further, based on the analysis of $\chi(t)$, we now present the property of $\zeta(t)$. Owing to the utilized projection operator, there exists a preassigned maxima of $\zeta(t)$ for $\forall t\in\left[t_{0},+\infty\right)$, denoted as $\zeta_{\max} > 0$. Therefore, for $\frac{1}{\zeta(t)}$, there consequently exists a minima such that:
		\begin{equation}
			\frac{1}{\zeta(t)} \ge \frac{1}{\zeta_{\max}} > 0
		\end{equation}
		This guarantees that $\frac{1}{\zeta(t)}$ is positive.
		
		Considering the time-derivative of $\frac{1}{\zeta(t)}$, by directly taking its time-derivative and combined with the compact form of $\dot{\zeta}(t)$, one can be obtained that:
		\begin{equation}\label{dzeta1}
			\begin{aligned}
				d(\frac{1}{\zeta(t)})/dt &= -\frac{2}{\zeta^{2}(t)}\cdot\lambda_{\zeta}\mu(t)\left[\zeta(t)-K_{\zeta}\chi(t)\zeta^{2}(t)\right]\\
				&= -2\lambda_{\zeta}\mu(t)\frac{1}{\zeta(t)} + 2\lambda_{\zeta}K_{\zeta}\mu(t)\chi(t)
			\end{aligned}
		\end{equation}
		
		Therefore, we integrating on both sides of equation (\ref{dzeta1}) from the initial time instant $t_{0}$ to the current time $t$, it can be obtained that:
		\begin{equation}\label{onezeta}
			\begin{aligned}
				\frac{1}{\zeta(t)} &= e^{-2\lambda_{\zeta}\int_{t_{0}}^{t}\mu(s)ds}\frac{1}{\zeta(t_{0})}\\
				&\quad + 2\int_{t_{0}}^{t}e^{-2\lambda_{\zeta}\int_{s}^{t}\mu(p)dp}\lambda_{\zeta}K_{\zeta}\mu(s)\chi(s)ds
			\end{aligned}
		\end{equation}
		Owing to the fact that $\chi(t)<1$ and $\mu(t)\in\left[0,1\right]$ holds (as we previously discussed in equation (\ref{chixiaoyuyi}) and (\ref{dotfzeta})), applying the comparison lemma yields:
		\begin{equation}\label{onedividezeta}
			\begin{aligned}
				\frac{1}{\zeta(t)} &\le e^{-2\lambda_{\zeta}(t-t_{0})}\frac{1}{\zeta(t_{0})}+ 2\lambda_{\zeta}K_{\zeta}\int_{t_{0}}^{t}e^{-2\lambda_{\zeta}(t-s)}ds\\
				&\le \frac{1}{\zeta(t_{0})}+K_{\zeta}
			\end{aligned}
		\end{equation} 
		Let the right side of equation (\ref{onedividezeta}) as $\zeta^{-1}_{\min}$. Correspondingly, for $\forall t\in\left[t_{0},+\infty\right)$, we have:
			$\zeta(t) \ge \frac{1}{\zeta^{-1}_{\min}} = \zeta_{\min} = \frac{1}{\frac{1}{\zeta(t_{0})}+K_{\zeta}}> 0$.

		This completes the proof of Corollary \ref{coroZeta}.
	\end{proof}
	
	\bibliographystyle{IEEEtran}
	\bibliography{CPC0528.bib}
	
\end{document}